\documentclass[draftclsnofoot, onecolumn]{IEEEtran}
\usepackage{ushort}
\usepackage[mathcal]{euscript}
\usepackage{amsmath}
\usepackage{amsthm}
\usepackage{amssymb}
\usepackage{graphicx}
\usepackage{algorithmic}
\usepackage{algorithm}
\usepackage{epsfig}
\usepackage{booktabs}% http://ctan.org/pkg/booktabs
\usepackage{colortbl}% http://ctan.org/pkg/colortbl
\usepackage{amsmath}% http://ctan.org/pkg/amsmath
\usepackage{xcolor}% http://ctan.org/pkg/xcolor
\usepackage{graphicx}% http://ctan.org/pkg/graphicx
\usepackage{cite} % aggregate refrences

\providecommand{\mbf}[1]{\mathbf{#1}}						 % Math boldface
\providecommand{\wt}[1]{\widetilde{#1}}					 % Wide tilde
\providecommand{\mbfwt}[1]{\wt{\mathbf{#1}}}				 % Math boldface with wide tilde
\providecommand{\mc}[1]{\mathcal{#1}}				 % Math Cal
\providecommand{\bsym}[1]{\boldsymbol{#1}}				 % Bold symbols by the bm package. Better than 																					 amsmath \pmb if the font exists.
\providecommand{\bsymwt}[1]{\widetilde{\boldsymbol{#1}}}	 % Bold symbol with wide tilde

\providecommand{\mbb}[1]{\mathbb{#1}}	

\colorlet{tableheadcolor}{gray!25} % Table header colour = 25% gray
\newcommand{\headcol}{\rowcolor{tableheadcolor}} %
\colorlet{tablerowcolor}{gray!10} % Table row separator colour = 10% gray
\newcommand{\rowcol}{\rowcolor{tablerowcolor}} %
\newcommand{\topline}{\arrayrulecolor{black}\specialrule{0.1em}{\abovetopsep}{0pt}%
            \arrayrulecolor{tableheadcolor}\specialrule{\belowrulesep}{0pt}{0pt}%
            \arrayrulecolor{black}}
\newcommand{\midline}{\arrayrulecolor{tableheadcolor}\specialrule{\aboverulesep}{0pt}{0pt}%
            \arrayrulecolor{black}\specialrule{\lightrulewidth}{0pt}{0pt}%
            \arrayrulecolor{white}\specialrule{\belowrulesep}{0pt}{0pt}%
            \arrayrulecolor{black}}
\DeclareMathOperator*{\argmin}{arg \, min}

\newtheorem{theorem}{Theorem}
\newtheorem{lemma}[theorem]{Lemma}

\newtheorem{property}[theorem]{Property}

\begin{document}

%\title{Finite Alphabet Constant Envelope QPSK MIMO Radar Waveform for Coexistence
%With Cellular Systems}
%\title{Finite Alphabet Constant Envelope QPSK Waveform for MIMO Radar with Spectrum Sharing Constraints}
\title{QPSK Waveform for MIMO Radar with Spectrum Sharing Constraints}
%\title{FACE QPSK Waveform Design for MIMO Radar}
%\title{On Finite Alphabet Constant Envelope QPSK Waveform for MIMO Radar}
%\author{\begin{tabular}{c}
% Awais Khawar, Ahmed Abdelhadi, and T. Charles Clancy \\
%\{awais, aabdelhadi, tcc\}@vt.edu \\
%Ted and Karyn Hume Center for National Security and Technology\\
%Bradley Department of Electrical and Computer Engineering\\
%Virginia Tech, Arlington, VA, 22203, USA
%\end{tabular}}
\author{Awais~Khawar,~\IEEEmembership{}Ahmed~Abdelhadi,~\IEEEmembership{}and~T. Charles~Clancy~\IEEEmembership{}

\thanks{Awais Khawar (awais@vt.edu) is with Virginia Polytechnic Institute and State University, Arlington, VA, 22203.

Ahmed Abdelhadi (aabdelhadi@vt.edu) is with Virginia Polytechnic Institute and State University, Arlington, VA, 22203.

T. Charles Clancy (tcc@vt.edu) is with Virginia Polytechnic Institute and State University, Arlington, VA, 22203.

This work was supported by DARPA under the SSPARC program. Contract Award Number: HR0011-14-C-0027. The views expressed are those of the author and do not reflect the official
policy or position of the Department of Defense or the U.S. Government.

Distribution Statement A: Approved for public release; distribution is unlimited.

}

}

\maketitle
%\thanks{We acknowledge the support from ONR.}

\begin{abstract}

%%%%%%%%%%%%%%%%%% 192/200 word abstract %%%%%%%%%%%%%%%%%%%%%%%%%

Multiple-input multiple-output (MIMO) radar is a relatively new concept in the field of radar signal processing. Many novel MIMO radar waveforms have been developed by considering various performance metrics and constraints. 
In this paper, we show that finite alphabet constant-envelope (FACE) quadrature-pulse shift keying (QPSK) waveforms can be designed to realize a given covariance matrix by transforming a constrained nonlinear optimization problem into an unconstrained nonlinear optimization problem. In addition, we design QPSK waveforms in a way that they don't cause interference to a cellular system, by steering nulls towards a selected base station (BS). The BS is selected according to our algorithm which guarantees minimum degradation in radar performance due to null space projection (NSP) of radar waveforms. We design QPSK waveforms with spectrum sharing constraints for a stationary and moving radar platform. We show that the waveform designed for stationary MIMO radar matches the desired beampattern closely, when the number of BS antennas $N^{\text{BS}}$ is considerably less than the number of radar antennas $M$, due to quasi-static interference channel. However, for moving radar the difference between designed and 
desired waveforms is larger than stationary radar, due to rapidly changing channel.

\end{abstract}

\begin{keywords}
MIMO Radar, Constant Envelope Waveform, QPSK, Spectrum Sharing
\end{keywords}

\section{Introduction}
%
%In the area of radar waveform design, waveforms are often designed to have certain properties, for example, low cross-correlation, clutter suppression, interference mitigation etc.  
%

An interesting concept for next generation of radars is multiple-input multiple-output (MIMO) radar systems; this has been an active area of research for the last couple of years \cite{LS08}. MIMO radars have been classified into widely-spaced \cite{HBC08}, where antenna elements are placed widely apart, and colocated \cite{LS07}, where antenna elements are placed next to each other. MIMO radars can transmit multiple signals, via its antenna elements, that can be different from each other, thus, resulting in waveform diversity. This gives MIMO radars an advantage over traditional phased-array radar systems which can only transmit scaled versions of single waveform and, thus, can't exploit waveform diversity.

Waveforms with constant-envelope (CE) are very desirable, in radar and communication system, from an implementation perspective, i.e., they allow power amplifiers to operate at or near saturation levels. CE waveforms are also popular due to their ability to be used with power efficient class C and class E power amplifiers and also with linear power amplifiers with no average power back-off into power amplifier. As a result, various researchers have proposed CE waveforms for communication systems; for example, CE multi-carrier modulation waveforms \cite{TS02}, such as CE orthogonal frequency division multiplexing (CE-OFDM) waveforms \cite{TAP+08}; and radar systems, for example, CE waveforms \cite{SLZ08}, CE binary-phase shift keying (CE-BPSK) waveforms \cite{ATPM11_FACE}, and CE quadrature-phase shift keying (CE-QPSK) waveforms \cite{SHC+14}.

Existing radar systems, depending upon their type and use, can be deployed any where between 3 MHz to 100 GHz of radio frequency (RF) spectrum. In this range, many of the bands are very desirable for international mobile telecommunication (IMT) purposes. For example, portions of the 700-3600 MHz band are in use by various second generation (2G), third generation (3G), and fourth generation (4G) cellular standards throughout the world. It is expected that mobile traffic volume will continue to increase as more and more devices will be connected to wireless networks. The current allocation of spectrum to wireless services is inadequate to support the growth in traffic volume. A solution to this spectrum congestion problem was presented in a report by President's Council of Advisers on Science and Technology (PCAST), which advocated to \textit{share} 1000 MHz of government-held spectrum \cite{PCAST12}. As a result, in the United States (U.S.), regulatory efforts are underway, by the Federal Communications 
Commission (FCC) along with the National Telecommunications and Information Administration (NTIA), to share government-held spectrum with commercial entities in the frequency band 3550-3650 MHz \cite{FCC12_SmallCells}. In the U.S., this frequency band is currently occupied by various services including radio navigation services by radars. The future of spectrum sharing in this band depends on novel interference mitigation methods to protect radars and commercial cellular systems from each others' interference \cite{SKA+14, Mo_radar, Awais_MILCOM}. Radar waveform design with interference mitigation properties is one way to address this problem, and this is the subject of this paper.

\subsection{Related Work}

Transmit beampattern design problem, to realize a given covariance matrix subject to various constraints, for MIMO radars is an active area of research; many researchers have proposed algorithms to solve this beampattern matching problem. Fuhrmann et al. proposed waveforms with arbitrary cross-correlation matrix by solving beampattern optimization problem, under the constant-modulus constraint, using various approaches \cite{FS08}. Aittomaki et al. proposed to solve beampattern optimization problem under the total power constraint as a least squares problem \cite{AK07}. Gong et al. proposed an optimal algorithm for omnidirectional beampattern design problem with the constraint to have sidelobes smaller than some predetermined threshold values \cite{GST14}. Hua et al. proposed transmit beampatterns with constraints on ripples, within the energy focusing section, and the transition bandwidth \cite{HA13}. However, many of the above approaches don't consider designing waveforms with finite alphabet and constant-
envelope property, which is very desirable from an implementation perspective. Ahmed et al. proposed a method to synthesize covariance matrix of BPSK waveforms with finite alphabet and constant-envelope property \cite{ATPM11_FACE}. They also proposed a similar solution for QPSK waveforms but it didn't satisfy the constant-envelope property. A method to synthesize covariance matrix of QPSK waveforms with finite alphabet and constant-envelope property was proposed by Sodagari et al. \cite{SHC+14}. However, they did not prove that such a method is possible. We prove the result in this paper and show that it is possible to synthesize covariance matrix of QPSK waveforms with finite alphabet and constant-envelope property.

As introduced earlier due to the congestion of frequency bands future communication systems will be deployed in radar bands. Thus, radars and communication systems are expected to share spectrum without causing interference to each other. For this purpose, radar waveforms should be designed in such a way that they not only mitigate interference to them but also mitigate interference by them to other systems \cite{SKC+12, KAC+14ICNC}. Transmit beampattern design by considering the spectrum sharing constraints is a fairly new problem. Sodagari et al. have proposed BPSK and QPSK transmit beampatterns by considering the constraint that the designed waveforms do not cause interference to a single communication system \cite{SHC+14}. This approach was extended to multiple communication systems, cellular system with multiple base stations, by Khawar et al. for BPSK transmit beampatterns \cite{KAC14DySPANWaveform, KAC+14DySPANProjection}. We extend this approach and consider optimizing QPSK transmit beampatterns for 
a cellular system with multiple base stations.

\subsection{Our Contributions}

In this paper, we make contributions in the areas of: 
%(a) finite alphabet constant-envelope QPSK waveform design and spectrum sharing between radars and cellular systems by shaping the newly designed QPSK radar waveform in a certain way. Our contributions are as follows:
\begin{itemize}
\item \textbf{Finite alphabet constant-envelope QPSK waveform:} In this area of MIMO radar waveform design, we make the following contribution: we prove that covariance matrix of finite alphabet constant-envelope QPSK waveform is positive semi-definite and the problem of designing waveform via solving a constrained optimization problem can be transformed into an un-constrained optimization problem.

\item \textbf{MIMO radar waveform with spectrum sharing constraints:} We design MIMO radar waveform for spectrum sharing with cellular systems. We modify the newly designed QPSK radar waveform in a way that it doesn't cause interference to communication system. We design QPSK waveform by considering the spectrum sharing constraints, i.e., the radar waveform should be designed in such a way that a cellular system experiences zero interference. We consider two cases: first, stationary maritime MIMO radar is considered which experiences a stationary or slowly moving interference channel. For this type of radar, waveform is designed by including the constraints in the unconstrained nonlinear optimization problem, due to the tractability of the constraints. Second, we consider a moving maritime MIMO radar which experiences interference channels that are fast enough not to be included in the optimization problem due to their intractability. For this type of radar, FACE QPSK waveform is designed which is then 
projected onto the null space of interference channel before transmission.

\end{itemize}

\subsection{Organization}
This paper is organized as follows. System model, which includes radar, communication system, interference channel, and cooperative RF environment model is discussed in Section \ref{sec:model}. Section \ref{sec:BP_prelim_FACE} introduces finite alphabet constant-envelope beampattern matching design problem. Section \ref{sec:QPSK_prelim} introduces QPSK radar waveforms and Section \ref{sec:QPSKproof} provides a proof of FACE QPSK waveform. Section \ref{sec:arch} discusses spectrum sharing architecture along with BS selection and projection algorithm. Section \ref{sec:nsp} designs QPSK waveforms with spectrum sharing constraints for stationary and moving radar platforms. Section \ref{sec:simulation}
discusses simulation setup and results. Section \ref{sec:conclude} concludes
the paper.

\renewcommand{\arraystretch}{1.5}
\begin{table}
\centering
\caption{Table of Notations}
\begin{tabular}{ll}
  \topline
  \headcol Notation &Description\\
  \midline
  		    $\mbfwt x(n)$			&Transmitted QPSK radar waveform\\
	\rowcol $\mbf a(\theta_k)$  	 &Steering vector to steer signal to 											  target angle $\theta_k$\\
			$\mbfwt r_k(n)$			 &Received radar waveform from target at $\theta_k$\\
 	\rowcol	$\mbfwt R$ 			 	&Correlation matrix of QPSK 											  		waveforms\\
			$\mbf s_j(n)$ 	 		&Signal transmitted by the 														$j^{\text{th}}$ UE in the 														$i^{\text{th}}$ cell\\
	\rowcol	$\mc L_i$ 				&Total number of user equipments 												(UEs) in the $i^{\text{th}}$ cell \\
			$\mc K$						&Total number of BSs \\
	\rowcol $M$ 						&Radar transmit/receive antennas \\
			$N_{\text{BS}}$				&BS transmit/receive antennas  \\ 
	\rowcol	$N_{\text{UE}}$				&UE transmit/receive antennas  \\
			$\mbf H_i$					&$i^{\text{th}}$ interference 															channel\\
	\rowcol $H_n$						&Hermite Polynomial\\
			$\mbf y_i(n)$				&Received signal at the 														$i^{\text{th}}$ BS\\
	\rowcol $\mbf P_i$					&Projection matrix for the 														$i^{\text{th}}$ channel \\
	
  \hline
\end{tabular}
\label{tab:Notations}
\end{table}

\subsection{Notations} Bold upper case letters, $\textbf{A}$, denote matrices while bold lower case letters, $\textbf{a}$, denote vectors. The $m^{\text{th}}$ column of matrix is denoted by $\textbf{a}_m$. For a matrix $\textbf{A}$, the conjugate and conjugate transposition are respectively denoted by $\textbf{A}^{\star}$ and $\textbf{A}^{H}$. The $m^{\text{th}}$ row and $n^{\text{th}}$ column element is denoted by $\textbf{A}(m,n)$. Real and complex, vectors and matrices are denoted by operators $\Re (\cdot)$ and $\Im (\cdot)$, respectively. A summary of notations is provided in Table \ref{tab:Notations}.

\section{System Model}\label{sec:model}

In this section, we introduce our system models for MIMO radar and cellular system. In addition, we introduce the cooperative RF sharing environment between radar and cellular system along with the definition of interference channel. 

\subsection{Radar Model}\label{sec:radar}

We consider waveform design for a colocated MIMO radar mounted on a ship. The radar has $M$ colocated transmit and receive antennas. The inter-element spacing between antenna elements is on the order of half the wavelength. The radars with colocated elements give better spatial resolution and target parameter estimation as compared to radars with widely spaced antenna elements \cite{LS07, HBC08}.

\subsection{Communication System}

We consider a MIMO cellular system, with $\mc K$ base stations, each equipped with $N_{\text{BS}}$ transmit and receive antennas, with the $i^{\text{th}}$ BS supporting $\mc L_i$ user equipments (UEs). Moreover, the UEs are also multi-antenna systems with $N_{\text{UE}}$ transmit and receive antennas. If $\mbf s_j(n)$ is the signals transmitted by the $j^{\text{th}}$ UE in the $i^{\text{th}}$ cell, then the received signal at the $i^{\text{th}}$ BS receiver can be written as
\begin{align}\label{eqn:RxPureComm}
\mbf y_i(n) = \sum_j \mbf H_{i,j} \: \mbf s_{j}(n) + \mbf w(n), \quad \, \, \text{for} \, \, 1 \leq i \leq \mc K \, \, \text{and} \, \, 1 \leq j \leq \mc L_i \nonumber
\end{align}
where $\mbf H_{i,j}$ is the channel matrix between the $i^{\text{th}}$ BS and the $j^{\text{th}}$ user and %$\mbf H_{N_R \times M_T}$ is the channel interference matrix between the radar and the communication system, 
$\mbf w(n)$ is the additive white Gaussian noise.

% H is RxAntenna x TxAntennas

\subsection{Interference Channel}

In our spectrum sharing model, radar shares $\mc K$ interference channels with cellular system. Let's define the $i^{\text{th}}$ interference channel as
\begin{equation}\label{Rx}
\mbf H_{i} \triangleq \begin{bmatrix} h_i^{(1,1)} &\cdots &h_i^{(1,M)}  
\\ \vdots  &\ddots &\vdots 
\\ h_i^{(N^{\text{BS}},1)}    &\cdots &h_i^{(N^{\text{BS}},M)} \end{bmatrix} \quad (N_{\text{BS}}  \times M)
\end{equation}
where $i=1, 2, \ldots, \mc K,$ and $h_i^{(l,k)}$ denotes the channel coefficient from the $k^{\text{th}}$ antenna element at the MIMO radar to the $l^{\text{th}}$ antenna element at the $i^{\text{th}}$ BS. We assume that elements of $\mbf H_i$ are independent, identically distributed (i.i.d.) and circularly symmetric complex Gaussian random variables with zero-mean and unit-variance, thus, having a i.i.d. Rayleigh distribution.

\subsection{Cooperative RF Environment}

Spectrum sharing between radars and communication systems can be envisioned in two types of RF environments, i.e., military radars sharing spectrum with military communication systems, we characterize it as \textit{Mil2Mil} sharing and military radars sharing spectrum with commercial communication systems, we characterize it as \textit{Mil2Com} sharing. In \textit{Mil2Mil} or \textit{Mil2Com} sharing, interference-channel state information (ICSI) can be provided to radars via feedback by military/commercial communication systems, if both systems are in a frequency division duplex (FDD) configuration \cite{TV05}. If both systems are in a time division duplex configuration, ICSI can be obtained via exploiting channel reciprocity \cite{TV05}. Regardless of the configuration of radars and communication systems, there is the incentive of zero interference, from radars, for communication systems if they collaborate in providing ICSI. Thus, we can safely assume the availability of ICSI for the sake of mitigating 
radar interference at communication systems.

%
%\section{Problem Formulation}\label{sec:Problem_formulation}
%
%We consider the problem of designing FACE QPSK waveform for MIMO radar when it is sharing spectrum with cellular system. We include the constraint that the radar waveform should be designed in such a way that it does not cause interference to cellular system. This problem is solved by introducing new constraints in the waveform design problem. Thus, the problem of beampattern design not only maximizes the received power at a number of target locations but also mitigates interference to cellular system.

\section{Finite Alphabet Constant-Envelope Beampattern Design}\label{sec:BP_prelim_FACE}
In this paper, we design QPSK waveforms having finite alphabets and constant-envelope property. We consider a uniform linear array (ULA) of $M$ transmit antennas with inter-element spacing of half-wavelength. Then, the transmitted QPSK signal is given as
\begin{equation}
\mbfwt x(n) = \begin{bmatrix}
\wt x_1(n) &\wt x_2(n) &\cdots &\wt x_M(n) 
\end{bmatrix}^T
\end{equation}
where $\wt x_m(n)$ is the QPSK signal from the $m^{\text{th}}$ transmit element at time index $n$. Then, the received signal from a target at location $\theta_k$ is given as
\begin{equation}\label{eqn:rxsignal}
\wt r_k(n) = \sum_{m=1}^{M}{e^{-j(m-1)\pi \sin\theta_k} \wt x_m(n)}, \quad k = 1,2, ...,K,
\end{equation}
where $K$ is the total number of targets. We can write the received signal compactly as
\begin{equation}
\wt r_k(n) = \mbf{a}^H(\theta_k)\mbfwt{x}(n)
\end{equation}
where $\mbf a(\theta_k)$ is the steering vector defined as
\begin{equation}
\mbf a(\theta_k) = \begin{bmatrix}
1 &e^{-j\pi \sin\theta_k}  &\cdots &e^{-j(M-1)\pi \sin\theta_k}\end{bmatrix}^T.
\end{equation}
We can write the power received at the target located at $\theta_k$ as
\begin{equation}
\begin{aligned}
P(\theta_k) & = \mathbb{E}\{\mbf{a}^H(\theta_k)\, \mbfwt{x}(n)\, \mbfwt{x}^H(n)\, \mbf{a}(\theta_k)\} \\
            & = \mbf{a}^H(\theta_k)\, \mbfwt{R} \, \mbf{a}(\theta_k)
\end{aligned}
\end{equation}
where $\mbfwt {R}$ is correlation matrix of the transmitted QPSK waveform. The desired QPSK beampattern $\phi(\theta_k)$ is formed by minimizing the square of the error between $P(\theta_k)$ and $\phi(\theta_k)$ through a cost function defined as 
\begin{equation}\label{eqn:obj_fn}
J(\mbfwt {R}) = \frac{1}{K}\sum_{k=1}^{K}{\Big(\mbf{a}^H(\theta_k)\, \mbfwt{R}\, \mbf{a}(\theta_k)-\phi(\theta_k)\Big)^2}.
\end{equation}
Since, $\mbfwt R$ is covariance matrix of the transmitted signal it must be positive semi-definite. Moreover, due to the interest in constant-envelope property of waveforms, all antennas must transmit at the same power level. The optimization problem in equation \eqref{eqn:obj_fn} has some constraints and, thus, can't be chosen freely. In order to design finite alphabet constant-envelope waveforms, we must satisfy the following constraints:
\begin{align*}
 C_1: \, &\mbf v^H \mbfwt{R}  \mbf v \geq 0, \qquad \qquad \; \; \; \forall \: \mbf v, \\
C_2: \, &\mbfwt{R} (m,m) = c, \qquad \qquad m=1,2, \ldots, M,
\end{align*}
where $C_1$ satisfies the `positive semi-definite' constraint and $C_2$ satisfies the `constant-envelope' constraint. Thus, we have a constrained nonlinear optimization problem given as
\begin{equation}\label{eqn:opt_const_smallR}
\begin{aligned}
& \underset{\mbfwt{R}}{\text{min}}
& & \frac{1}{K}\sum_{k=1}^{K}{\Big(\mbf{a}^H(\theta_k) \, \mbfwt{R} \, \mbf{a}(\theta_k)-\phi(\theta_k)\Big)^2} \\
& \text{subject to}
& & \mbf{v}^H\mbfwt{R}\mbf v \geq 0, \;\;\;\;\;\;\;\; \forall \:\mbf{v},\\
& & &  \mbfwt{R}(m,m) = c, \;\;\;\;\; m = 1,2, ...,M.
\end{aligned}
\end{equation}
Ahmed et al. showed that, by using multi-dimensional spherical coordinates, this constrained nonlinear optimization can be transformed into an unconstrained nonlinear optimization \cite{ATPM11_UnconstSynthesis}. Once $\mbfwt{R} $ is synthesized, the waveform matrix $\mbfwt X$ with $N$ samples is given as
\begin{equation}
\mbfwt X = \begin{bmatrix}
\mbfwt x(1) &\mbfwt x(2) &\cdots &\mbfwt x(N)
\end{bmatrix}^T.
\end{equation}
This can be realized from
\begin{equation}
\mbfwt X = \bsym{\mathcal{X}} \bsym \Lambda^{1/2} \mbf W^H \label{eqn:GaussX}
\end{equation}  
where  $\bsym{\mathcal{X}} \in \mathcal{C}^{N \times M}$ is a matrix of zero mean and unit variance Gaussian random variables, $\bsym \Lambda \in \mathcal{R}^{M \times M}$ is the diagonal matrix of eigenvalues, and $\mbf W \in \mathcal{C}^{M \times M}$ is the matrix of eigenvectors of $\mbfwt{R} $ \cite{HKO01}. Note that $\mbfwt X$ has Gaussian distribution due to $\bsym{\mathcal{X}}$ but the waveform produced is not guaranteed to have the CE property. 
%In the following sections we design FACE QPSK waveforms

\section{Finite Alphabet Constant-Envelope QPSK Waveforms}\label{sec:QPSK_prelim}

In \cite{SHC+14}, an algorithm to synthesize FACE QPSK waveforms to realize a given covariance matrix, $\mbfwt R$, with complex entries was presented. However, it was not proved that such a covariance matrix is positive semi-definite and the constrained nonlinear optimization problem can be transformed into an un-constrained nonlinear optimization problem, we prove the claim in this paper.

Consider zero mean and unit variance Gaussian random variables (RVs) $\wt x_m$ and $\wt y_m$ that can be mapped onto a QPSK RV $\wt z_m$ through, as in \cite{SHC+14},
\begin{equation}\label{eqn:QPSK_RV}
\wt z_m = \frac{1}{\sqrt{2}}\bigg[ \text{sign}(\wt x_m) + \jmath \, \text{sign} (\wt y_m) \bigg].
\end{equation}
Then, it is straight forward to write the $(p,q)$th element of the complex covariance matrix as
\begin{equation}
\mbb E \{\wt z_p \wt z_q\} = \gamma_{pq} = \gamma_{\Re_{pq}} + \jmath \, \gamma_{\Im_{pq}}
\end{equation}
where $\gamma_{\Re_{pq}}$ and $\gamma_{\Im_{pq}}$ are the real and imaginary parts of $\gamma_{{pq}}$, respectively. If, Gaussian RVs $\wt x_p, \wt x_q, \wt y_p$, and $\wt y_q$ are chosen such that  
\begin{align}\label{eqn:ExpProperty}
\mbb E \{\wt x_p \wt x_q\} &= \mbb E \{\wt y_p \wt y_q\} \nonumber \\
\mbb E \{\wt x_p \wt y_q\} &= - \mbb E \{\wt y_p \wt x_q\} 
\end{align}
then we can write the real and imaginary parts of $\gamma_{pq}$ as
\begin{align}
\gamma_{\Re_{pq}} &= \mbb E \Big\{ \text{sign}(\wt x_p) \text{sign} (\wt x_q) \Big\} \nonumber \\
\gamma_{\Im_{pq}} &= \mbb E \Big\{ \text{sign}(\wt y_p)\text{sign} (\wt x_q) \Big\}\cdot
\end{align}
Then, from equation \eqref{eqn:ProofXcor} Appendix \ref{sec:ProofGaussian}, we have
\begin{equation}\label{eqn:ExpectZpZq}
\mbb E \{\wt z_p \wt z_q\} = \frac{2}{\pi} \Bigg[ \sin^{-1} \bigg(\mbb E \{\wt x_p \wt x_q\} \bigg) + \jmath \, \sin^{-1} \bigg(\mbb E \{\wt y_p \wt x_q\} \bigg) \Bigg].
\end{equation}
The complex Gaussian covariance matrix $\mbfwt  R_g$ is defined as
\begin{equation}
\mbfwt R_g \triangleq \Re ( \mbf  R_g  ) + \jmath \, \Im (\mbf R_g ) \label{eq:def_Rg}
\end{equation}
where $\Re ( \mbf  R_g  )$ and $\Im (\mbf R_g )$ both have real entries, since $\mbf  R_g $ is a real Gaussian covariance matrix.
% and generated using equation \eqref{}. 
Then, equation \eqref{eqn:ExpectZpZq} can be written as
\begin{equation}
\mbfwt  R = \frac{2}{\pi} \bigg[ \sin^{-1} \Big( \Re ( \mbf  R_g ) \Big) + \jmath \, \sin^{-1} \Big( \Im ( \mbf  R_g ) \Big) \bigg]. \label{eqn:def_R}
\end{equation}
In \cite{SHC+14}, it is proposed to construct complex Gaussian covariance matrix via transform $\mbfwt R_g = \mbfwt U^H \mbfwt U$, where $\mbfwt U$ is given by equation \eqref{eqn:Utilda}. Then, $\mbfwt U$ can be written as 
\begin{equation}
\mbfwt U = \Re (\mbfwt U) + \jmath \Im (\mbfwt U)
\end{equation}
where $\Re (\mbfwt U)$ and $\Im (\mbfwt U)$ are given by equations \eqref{eqn:realR} and \eqref{eqn:imagR}, respectively. Alternately, $\mbfwt R_g$ can also be expressed as
\begin{align}\label{eqn:RgDefined}
\mbfwt R_g = \bigg[\Re (\mbfwt U)^H \Re (\mbfwt U) + \Im (\mbfwt U)^H \Im (\mbfwt U)\bigg] +\jmath \, \bigg[ \Re (\mbfwt U)^H\Im (\mbfwt U)-\Im (\mbfwt U)^H\Re (\mbfwt U)\bigg].
\end{align}

%\begin{figure*}
\begin{equation}\label{eqn:Utilda}
\mbfwt U = 
\begin{pmatrix}
  e^{j \psi_1}       &e^{j \psi_2} \sin(\psi_{21}) & e^{j \psi_3}\sin(\psi_{31})\sin(\psi_{32}) & \cdots  & e^{j \psi_M}\prod_{m=1}^{M-1} \sin(\psi_{Mm})                      \\
  0       &e^{j \psi_2} \cos(\psi_{21}) &e^{j \psi_3} \sin(\psi_{31})\cos(\psi_{32}) & \cdots  &e^{j \psi_M} \prod_{m=1}^{M-2}  \sin(\psi_{Mm})\cos(\psi_{M,M-1})   \\
  0       &         0     &               e^{j \psi_3} \cos(\psi_{31}) & \ddots  & \vdots                                                 \\
  \vdots  &      \vdots     &             \ddots             & \cdots  &e^{j \psi_M} \sin(\psi_{M1})\cos(\psi_{M2})                         \\
  0       &         0       &             \cdots             & \cdots  &e^{j \psi_M} \cos(\psi_{M1})
\end{pmatrix}
\end{equation}
%\hrulefill
%\vspace*{4pt}
%\end{figure*}

%\begin{figure*}
\begin{equation}\label{eqn:realR}
\Re \big(\mbfwt U \big) = 
\begin{pmatrix}
  \cos(\psi_1)       &\cos(\psi_2) \sin(\psi_{21}) &\cos(\psi_3) \sin(\psi_{31})\sin(\psi_{32}) &\cdots  &\cos(\psi_M)  \prod_{m=1}^{M-1} \sin(\psi_{Mm})                      \\
  0       &\cos(\psi_2) \cos(\psi_{21}) &\cos(\psi_3) \sin(\psi_{31})\cos(\psi_{32}) & \cdots  &\cos(\psi_M) \prod_{m=1}^{M-2}  \sin(\psi_{Mm})\cos(\psi_{M,M-1})   \\
  0       &         0     &\cos(\psi_3)                \cos(\psi_{31}) & \ddots  & \vdots                                                 \\
  \vdots  &      \vdots     &             \ddots             & \cdots  &\cos(\psi_M) \sin(\psi_{M1})\cos(\psi_{M2})                         \\
  0       &         0       &             \cdots             & \cdots  &\cos(\psi_M) \cos(\psi_{M1})
\end{pmatrix}
\end{equation}
%\hrulefill
%\vspace*{4pt}
%\end{figure*}

%\begin{figure*}
\begin{equation}\label{eqn:imagR}
\Im \big(\mbfwt U \big) = 
\begin{pmatrix}
  \sin(\psi_1)       &\sin(\psi_2) \sin(\psi_{21}) &\sin(\psi_3) \sin(\psi_{31})\sin(\psi_{32}) &\cdots  &\sin(\psi_M)  \prod_{m=1}^{M-1} \sin(\psi_{Mm})                      \\
  0       &\sin(\psi_2) \cos(\psi_{21}) &\sin(\psi_3) \sin(\psi_{31})\cos(\psi_{32}) & \cdots  &\sin(\psi_M) \prod_{m=1}^{M-2}  \sin(\psi_{Mm})\cos(\psi_{M,M-1})   \\
  0       &         0     &\sin(\psi_3)                \cos(\psi_{31}) & \ddots  & \vdots                                                 \\
  \vdots  &      \vdots     &             \ddots             & \cdots  &\sin(\psi_M) \sin(\psi_{M1})\cos(\psi_{M2})                         \\
  0       &         0       &             \cdots             & \cdots  &\sin(\psi_M) \cos(\psi_{M1})
\end{pmatrix}
\end{equation}
%\hrulefill
%\vspace*{4pt}
%\end{figure*}

\setcounter{theorem}{0}

\begin{lemma} \label{L1}
If $\mbf R_g$ is a covariance matrix and 
\begin{equation}
\mbfwt  R_g = \Re ( \mbf  R_g  ) + \jmath \, \Im (\mbf R_g ) 
\end{equation}
then the complex covariance matrix $\mbfwt R_g$ will always be positive semi-definite.
\end{lemma}
\begin{proof}
Please see Appendix \ref{app:Proofs}.
\end{proof}
%
%The relationship between complex Gaussian covariance matrix and covariance matrix of QPSK waveforms can be written as
%\begin{equation}
%\mbfwt  R = \frac{2}{\pi} \left[ \sin^{-1} \left( \Re ( \mbf  R_g ) \right) + \jmath \, \sin^{-1} \left( \Im ( \mbf  R_g ) \right) \right] \label{eq:def_R}
%\end{equation}
Lemma \ref{L1} satisfies constraint $C_1$ and $\mbfwt R_g$ also satisfies constraint $C_2$ for $c=1$. This helps to transform constrained nonlinear optimization into unconstrained nonlinear optimization in the following section. 

In order to generate QPSK waveforms we define $N \times 2M$ matrix $\mbfwt S$, of Gaussian RVs, as
\begin{equation}
\mbfwt S \triangleq \begin{bmatrix}\mbfwt X &\mbfwt Y \end{bmatrix}
\end{equation}
where $\mbfwt X$ and $\mbfwt Y$ are of each size $N \times M$, representing real and imaginary parts of QPSK waveform matrix, which is given as
\begin{equation}
\mbfwt Z = \frac{1}{\sqrt{2}}\bigg[ \text{sign}(\mbfwt X) + \jmath \, \text{sign} (\mbfwt Y) \bigg].
\end{equation}
The covariance matrix of $\mbfwt S$ is given as
\begin{equation}
\mbfwt R_{\mbfwt S} = \mbb E \{\mbfwt S^H \mbfwt S \} = \begin{bmatrix} \Re ( \mbf  R_g ) &\Im ( \mbf  R_g ) \\
-\Im ( \mbf  R_g ) &\Re ( \mbf  R_g ) \end{bmatrix}\cdot
\end{equation}
QPSK waveform matrix $\mbfwt Z$ can be realized by the matrix $\mbfwt S$ of Gaussian RVs which can be generated using equation \eqref{eqn:GaussX} by utilizing $\mbfwt R_{\mbfwt S}$.

\section{Gaussian Covariance Matrix Synthesis for Desired QPSK Beampattern}\label{sec:QPSKproof}

In this section, we prove that the desired QPSK beampattern can be directly synthesized by using the complex covariance matrix, $\mbfwt R_g$, for complex Gaussian RVs. This generates $M$ QPSK waveforms for the desired beampattern which satisfy the property of finite alphabet and constant-envelope.  By exploiting the relationship between the complex Gaussian RVs and QPSK RVs we have
% in equation \eqref{} one can write
\begin{equation}\label{eqn:def_RSecV}
\mbfwt  R = \frac{2}{\pi} \Bigg[ \sin^{-1} \bigg( \Re ( \mbf  R_g ) \bigg) + \jmath \, \sin^{-1} \bigg( \Im ( \mbf  R_g ) \bigg) \Bigg]. 
\end{equation}

\begin{lemma} \label{L2}
If $\mbfwt R_g$ is a complex covariance matrix and
\begin{equation*}
\mbfwt  R = \frac{2}{\pi} \Bigg[ \sin^{-1} \bigg( \Re ( \mbf  R_g ) \bigg) + \jmath \, \sin^{-1} \bigg( \Im ( \mbf  R_g ) \bigg) \Bigg] 
\end{equation*}
then $\mbfwt R$ will always be positive semi-definite.
\end{lemma}
\begin{proof}
Please see Appendix \ref{app:Proofs}.
\end{proof}

Using equation \eqref{eqn:def_RSecV} we can rewrite the optimization problem in equation \eqref{eqn:opt_const_smallR} as
\begin{equation}\label{eqn:opt_unct_smallR}
\begin{aligned}
& \underset{\mbfwt{R}}{\text{min}}
& &  \frac{1}{K} \sum_{k=1}^K \Bigg[ \frac{2}{\pi} \mbf a^H(\theta_k) \bigg\{\sin^{-1} \bigg(\Re (\mbf R_g)\bigg) + \jmath \sin^{-1} \bigg(\Im (\mbf R_g)\bigg) \bigg\} \mbf{a}(\theta_k)-\phi(\theta_k)\Bigg]^2 \\
& \text{subject to}
& & \mbf{v}^H\mbfwt{R}\mbf v \geq 0, \;\;\;\;\;\;\;\; \forall \:\mbf{v},\\
& & &  \mbfwt{R}(m,m) = c, \;\;\;\;\; m = 1,2, ...,M.
\end{aligned}
\end{equation}
%\begin{align}
%J (\mbf R_g) &= \frac{1}{K} \sum_{k=1}^K \Bigg[ \frac{2}{\pi} \mbf a^H(\theta_k) \bigg\{\sin^{-1} \bigg(\Re (\mbf R_g)\bigg) \nonumber \\
%&+ \jmath \sin^{-1} \bigg(\Im (\mbf R_g)\bigg) \bigg\} \mbf a(\theta_k) - \alpha \phi(\theta_k) \Bigg]^2
%\end{align}
%with the new constraints
%.......
%
%
%\begin{equation}\label{eqn:opt_ct}
%\begin{aligned}
%& \underset{\mbf{R}}{\text{min}}
%& & \frac{1}{K}\sum_{k=1}^{K}{\Big(\mbf{a}^H(\theta_k) \, \mbf{R} \, \mbf{a}(\theta_k)-\phi(\theta_k)\Big)^2} \\
%& \text{subject to}
%& & \mbf{v}^H\mbf{R}\mbf{v}, \;\;\;\;\;\;\;\;\;\;\;\;\;\;\;\; \forall \:\mbf{v}\\
%& & &  \mbf{R}(m,m) = c, \;\;\;\;\; m = 1,2, ...,M.
%\end{aligned}
%\end{equation}

%\begin{figure*}[t]
\begin{align}\label{eqn:JthetaDetailed}
J (\bsym \Theta) = \frac{1}{K} \sum_{k=1}^K \Bigg[ \frac{2}{\pi} \mbf a^H(\theta_k) &\bigg\{\sin^{-1} \bigg(\Re (\mbfwt U)^H \Re (\mbfwt U)+ \Im (\mbfwt U)^H \Im (\mbfwt U)\bigg) \nonumber \\
&+\jmath \sin^{-1} \bigg( \Re (\mbfwt U)^H\Im (\mbfwt U)-\Im (\mbfwt U)^H\Re (\mbfwt U)\bigg) \bigg\}\mbf a^H(\theta_k) - \alpha \phi(\theta_k) \Bigg]^2
\end{align}
%\end{figure*}

Since, the matrix $\mbfwt U$ is already known, we can formulate $\mbfwt R_g$ via equation \eqref{eqn:RgDefined}. We can also write the $(p,q)$th element of the upper triangular matrix $\mbfwt R_g$ by first writing the $(p,q)$th element of the upper triangular matrix $\Re \big(\mbf R_g(p,q)\big)$ as
\begin{align}
\Re &\big(\mbf R_g(p,q)\big) = \begin{cases}
\prod_{l=1}^{q-1} \sin (\Psi_{ql}) \prod_{s=1}^{p} \prod_{u=1}^{q} f(s,u), \, &p > q 
\\1, &p=q
\end{cases}
\end{align}
where $f(s,u) = \cos (\Psi_{s}) \cos (\Psi_{u})   + \sin (\Psi_{s}) \sin (\Psi_{u})$; and the $(p,q)$th element of the upper triangular matrix $\Im \big(\mbf R_g(p,q)\big)$ as
\begin{align}
\Im \big(\mbf R_g(p,q)\big) &= \begin{cases}
 g(p,q) \prod_{l=1}^{q-1} \sin (\Psi_{ql}), &p > q 
\\0, &p=q
\end{cases}
\end{align}
where $g(p,q) = \cos (\Psi_{p})  \sin (\Psi_{q}) + \sin (\Psi_{p}) \cos (\Psi_{q}) $. Thus, we can write the $(p,q)$th element of the upper triangular matrix $\mbfwt R_g$ as
%Note that by construction $\Re \big(\mbf R_g(p,q)\big)$ is symmetric and $\Im \big(\mbf R_g(p,q)\big)$ is skew-symmetric. 
\begin{equation}
\mbfwt R_g(p,q) = 
\begin{cases}
\Re \big(\mbf R_g(p,q)\big) + \jmath \Im \big(\mbf R_g(p,q)\big), \, &p > q 
\\1, &p=q.
\end{cases}
\end{equation}

By utilizing the information of $\mbfwt U$, the constrained optimization problem in equation \eqref{eqn:opt_unct_smallR} can be transformed into an unconstrained optimization problem that can be written as equation \eqref{eqn:JthetaDetailed},
%\begin{align}\label{eqn:JthetaDetailed}
%&J (\bsym \Theta) \nonumber \\
%&= \frac{1}{K} \sum_{k=1}^K \Bigg[ \frac{2}{\pi} \mbf a^H(\theta_k) \bigg\{\sin^{-1} \bigg(\Re (\mbfwt U)^H \Re (\mbfwt U)+ \Im (\mbfwt U)^H \Im (\mbfwt U)\bigg)  \nonumber \\
%&+ \jmath \sin^{-1} \bigg( \Re (\mbfwt U)^H\Im (\mbfwt U)-\Im (\mbfwt U)^H\Re (\mbfwt U)\bigg) \bigg\} \mbf a^H(\theta_k) - \alpha \phi(\theta_k) \Bigg]^2
%\end{align}
where 
\begin{equation}
\bsym \Theta = \begin{bmatrix} \bsym \Psi^T &\bsymwt \Psi^T &\alpha
\end{bmatrix}^T, 
\end{equation}
and
\begin{align*}
\bsym \Psi^T  &= \begin{bmatrix} \Psi_{21} &\Psi_{21} &\cdots &\Psi_{21} \end{bmatrix}^T, \\
\bsymwt \Psi^T  &= \begin{bmatrix}
\Psi_{1} &\Psi_{2} &\cdots &\Psi_{M} \end{bmatrix}^T.
\end{align*}
The optimization is over $M(M-1)/2 +M$ elements $\Psi_{mn}$ and $\Psi_{l}$. The advantage of this approach lies in the free selection of elements of $\bsym \Theta$ without effecting the positive semi-definite property and diagonal elements of $\mbfwt R_g$. Noting that $\mbfwt U$ and $\mbfwt R_g$ are functions of $\bsym \Theta$, we can alternately write the cost-function, in equation \eqref{eqn:JthetaDetailed}, as 
%\begin{align}
%J (\bsym \Theta) &= \frac{1}{K} \sum_{k=1}^K \Bigg[ \frac{2}{\pi} \mbf a^H(\theta_k) \bigg\{\sin^{-1} \big(\Re (\mbf R_g)\big) \nonumber \\
%&+ \jmath \sin^{-1} \big(\Im (\mbf R_g)\big) \bigg\} \mbf a(\theta_k) - \alpha \phi(\theta_k) \Bigg]^2
%\end{align}
\begin{align}\label{eqn:35}
J (\bsym \Theta) &= \frac{1}{K} \sum_{k=1}^K \bigg[ \frac{2}{\pi} \mbf a^H(\theta_k) \sin^{-1} \Big(\Re (\mbf R_g)\Big) \mbf a(\theta_k) + \frac{2\jmath}{\pi} \mbf a^H(\theta_k)  \sin^{-1} \Big(\Im (\mbf R_g)\Big) \mbf a(\theta_k) - \alpha \phi(\theta_k) \bigg]^2\cdot 
\end{align}
First, the partial differentiation of $J (\bsym \Theta)$ with respect to any element of $\bsym \Psi$, say $\Psi_{mn}$, can be found as
\begin{align}\label{eqn:partialJ}
\frac{\partial J (\bsym \Theta) }{\partial \Psi_{mn}} =& \Bigg[ \frac{2}{K} \sum_{k=1}^K \bigg\{ \frac{2}{\pi} \mbf a^H(\theta_k) \sin^{-1} \big(\Re (\mbf R_g)\big) \mbf a(\theta_k) + \frac{2\jmath}{\pi} \mbf a^H(\theta_k)  \sin^{-1} \big(\Im (\mbf R_g)\big) \mbf a(\theta_k) - \alpha \phi(\theta_k)\bigg\} \Bigg] \nonumber \\
&\times \Bigg[ \frac{\partial}{\partial \Psi_{mn}}\bigg\{ \frac{2}{\pi} \mbf a^H(\theta_k) \sin^{-1} \big(\Re (\mbf R_g)\big) \mbf a(\theta_k) + \frac{2\jmath}{\pi} \mbf a^H(\theta_k)  \sin^{-1} \big(\Im (\mbf R_g)\big) \mbf a(\theta_k)\bigg\} \Bigg]\cdot
\end{align}

The matrix $\Re (\mbf R_g)$ is real and symmetric, i.e., $\Re \big(\mbf R_g(p,q)\big)=\Re \big(\mbf R_g(q,p)\big)$, at the same time, $\Im (\mbf R_g)$ has real entries but is skew-symmetric, i.e., $\Im \big(\mbf R_g(p,q)\big)=-\Im \big(\mbf R_g(q,p)\big)$. These observations
%, along with the result of Lemma \ref{L3}, 
enables us to write equation \eqref{eqn:partialJ} in a simpler form
%\begin{align}\label{eqn:1stSimplification}
% &\frac{\partial J (\bsym \Theta)}{\partial \Psi_{mn}} = \Bigg[ \frac{4}{K} \sum_{k=1}^K \bigg\{ \frac{2}{\pi} \mbf a^H(\theta_k) \sin^{-1} \big(\Re (\mbf R_g)\big) \mbf a(\theta_k) \nonumber \\ &+ \frac{2\jmath}{\pi} \mbf a^H(\theta_k)  \sin^{-1} \big(\Im (\mbf R_g)\big) \mbf a(\theta_k) - \alpha \phi(\theta_k)\bigg\} \Bigg] \nonumber \\
%&\times \Bigg[\bigg\{ \frac{2}{\pi}  \sum_{p=1}^{M-1} \sum_{q=p+1}^{M} \frac{\cos\big(\pi|p-q|\sin(\theta_k)\big)}{\sqrt{1-\Re \big(\mbf R_g^2(p,q)\big)}} \frac{\partial \Re \big(\mbf R_g(p,q)\big)}{\partial \Psi_{mn}} \bigg\} \nonumber \\ &+ \bigg\{ \frac{\jmath}{\pi} \sum_{p=1}^{M-1} \sum_{q=p+1}^{M} \frac{\cos\big(\pi|p-q|\sin(\theta_k)\big)}{\sqrt{1-\Im \big(\mbf R_g^2(p,q)\big)}} \frac{\partial \Im \big(\mbf R_g(p,q)\big)}{\partial \Psi_{mn}}\nonumber \\
%&- \sum_{p=1}^{M-1} \sum_{q=p+1}^{M} \frac{\cos\big(\pi|p-q|\sin(\theta_k)\big)}{\sqrt{1+\Im \big(\mbf R_g^2(p,q)\big)}} \frac{\partial \Im \big(\mbf R_g(p,q)\big)}{\partial \Psi_{mn}} \bigg\} \Bigg]
%\end{align}
\begin{align}\label{eqn:1stSimplification1stElement}
\frac{\partial J (\bsym \Theta)}{\partial \Psi_{mn}} =& \Bigg[ \frac{4}{K} \sum_{k=1}^K \bigg\{ \frac{2}{\pi} \mbf a^H(\theta_k) \sin^{-1} \big(\Re (\mbf R_g)\big) \mbf a(\theta_k) + \frac{2\jmath}{\pi} \mbf a^H(\theta_k)  \sin^{-1} \big(\Im (\mbf R_g)\big) \mbf a(\theta_k) - \alpha \phi(\theta_k)\bigg\} \Bigg] \nonumber \\
&\times \Bigg[\frac{2}{\pi}  \sum_{p=1}^{M-1} \sum_{q=p+1}^{M} \frac{\cos\big(\pi|p-q|\sin(\theta_k)\big)}{\sqrt{1-\Re \big(\mbf R_g^2(p,q)\big)}} \frac{\partial \Re \big(\mbf R_g(p,q)\big)}{\partial \Psi_{mn}} \Bigg]\cdot
\end{align}
Moreover, $\Re (\mbf R_g)$ contains only $(M-1)$ terms which depend on $\Psi_{mn}$, thus, equation \eqref{eqn:1stSimplification1stElement} further simplifies as
\begin{align}\label{eqn:2ndSimplification1stElement}
\frac{\partial J (\bsym \Theta)}{\partial \Psi_{mn}} =& \frac{8}{\pi K} \Bigg[  \sum_{k=1}^K \bigg\{ \frac{2}{\pi} \mbf a^H(\theta_k) \sin^{-1} \big(\Re (\mbf R_g)\big) \mbf a(\theta_k) + \frac{2\jmath}{\pi} \mbf a^H(\theta_k)  \sin^{-1} \big(\Im (\mbf R_g)\big) \mbf a(\theta_k) - \alpha \phi(\theta_k)\bigg\} \Bigg] \nonumber \\
&\times \Bigg[\bigg\{ \sum_{p=1}^{m-1} \frac{\cos\big(\pi|p-m|\sin(\theta_k)\big)}{\sqrt{1-\Re \big(\mbf R_g^2(p,m)\big)}} \frac{\partial \Re \big(\mbf R_g(p,m)\big)}{\partial \Psi_{mn}} + \sum_{q=m+1}^{M} \frac{\cos\big(\pi|m-q|\sin(\theta_k)\big)}{\sqrt{1-\Re \big(\mbf R_g^2(m,q)\big)}} \frac{\partial \Re \big(\mbf R_g(m,q)\big)}{\partial \Psi_{mn}} \bigg\} \Bigg].
\end{align}
Second, the partial differentiation of $J (\bsym \Theta)$ with respect to any element of $\bsymwt \Psi$, say $\Psi_l$, can be found in the same manner as was found for $\Psi_{mn}$, i.e.,
\begin{align}\label{eqn:2ndElement}
\frac{\partial J (\bsym \Theta)}{\partial \Psi_{l}} =& \frac{8}{\pi K} \Bigg[ \sum_{k=1}^K \bigg\{ \frac{2}{\pi} \mbf a^H(\theta_k) \sin^{-1} \big(\Re (\mbf R_g)\big) \mbf a(\theta_k) + \frac{2\jmath}{\pi} \mbf a^H(\theta_k)  \sin^{-1} \big(\Im (\mbf R_g)\big) \mbf a(\theta_k) - \alpha \phi(\theta_k)\bigg\} \Bigg] \nonumber \\
&\times \Bigg[  \sum_{p=1}^{M-1} \sum_{q=p+1}^{M} \frac{\cos\big(\pi|p-q|\sin(\theta_k)\big)}{\sqrt{1-\Re \big(\mbf R_g^2(p,q)\big)}} \frac{\partial \Re \big(\mbf R_g(p,q)\big)}{\partial \Psi_{l}} \Bigg]\cdot
\end{align}
Finally, the partial differentiation of $J (\bsym \Theta)$ with respect to $\alpha$ is
\begin{align}%\label{eqn:2ndElement}
\frac{\partial J (\bsym \Theta)}{\partial \alpha} =\frac{-2 \phi(\theta_k)}{K} \Bigg[ \sum_{k=1}^K \bigg\{ \frac{2}{\pi} \mbf a^H(\theta_k) \sin^{-1} \big(\Re (\mbf R_g)\big) \mbf a(\theta_k) + \frac{2\jmath}{\pi} \mbf a^H(\theta_k)  \sin^{-1} \big(\Im (\mbf R_g)\big) \mbf a(\theta_k) - \alpha \phi(\theta_k)\bigg\} \Bigg]. 
\end{align}

\section{Radar-Cellular System Spectrum Sharing}\label{sec:arch}

%After introducing our radar and cellular system models we can now discuss the spectrum sharing scenario between radar and cellular system. 

%In our sharing architecture  are the co-primary users of 

%The 

In the following sections, we will discuss our spectrum sharing architecture and spectrum sharing algorithms for the 3550-3650 MHz band under consideration, which is co-shared by MIMO radar and cellular systems. .

\begin{figure}
\centering
	\includegraphics[width=\linewidth]{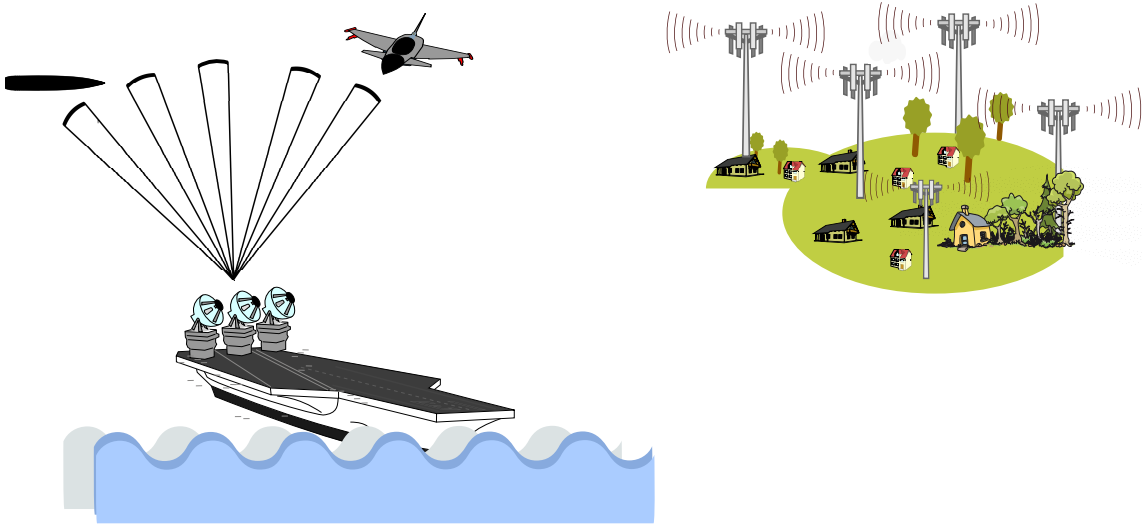} 
	\caption{Spectrum Sharing Scenario: Seaborne MIMO radar sharing spectrum with a cellular system.}
		\label{fig:scenario}
\end{figure}

\subsection{Architecture}
Considering the coexistence scenario in Fig. \ref{fig:scenario}, where the radar is sharing $\mc K$ interference channels with the cellular system, the received signal at the $i^{\text{th}}$ BS can be written as
\begin{equation}\label{eqn:RxRadarComm}
\mbf y_i(n) = \mbf H_i \mbfwt x(n) + \sum_j \mbf H_{i,j} \: \mbf s_{j}(n) + \mbf w(n)
\end{equation}
In order to avoid interference to the $i^{\text{th}}$ BS, the radar shapes its waveform $\mbfwt x(n)$ such that it is in the null-space of $\mbf H_i$, i.e. $\mbf H_i \mbfwt x(n) = \bsym 0$.
% in order to avoid interference to the $i^{\text{th}}$ BS, i.e., $\mbf H_i \mbf x(t) = \bsym 0$ so that $\mbf r_i(t)$ has equation \eqref{eqn:RxPureComm} instead of equation \eqref{eqn:RxRadarComm}. 

%The null space projection is performed using Algorithm \eqref{alg:M2C_Radar}.

\subsection{Projection Matrix}

In this section, we formulate a projection algorithm to project the radar signal onto the null space of interference channel $\mbf H_i$. Assuming, the MIMO radar has ICSI for all $\mbf H_i$ interference channels, either through feedback or channel reciprocity, we can perform a singular value decomposition (SVD) to find the null space of $\mbf H_i$ and use it to construct a projector matrix. First, we find SVD of $\mbf H_i$, i.e., 
\begin{equation}
\mbf H_i = \mbf U_i \bsym \Sigma_i \mbf V_i^H.
\end{equation}
Now, let us define 
\begin{equation}
\bsymwt \Sigma_i \triangleq \text{diag} (\wt \sigma_{i,1}, \wt \sigma_{i,2}, \ldots, \wt \sigma_{i,p})
\end{equation}
where $p \triangleq \min (N^{\text{BS}},M)$ and 
$\wt \sigma_{i,1} > \wt \sigma_{i,2} > \cdots > \wt \sigma_{i,q} > \wt \sigma_{i,q+1} = \wt \sigma_{i,q+2} = \cdots \wt \sigma_{i,p} = 0$. Next, we define
\begin{equation}
\bsymwt {\Sigma}_i^\prime \triangleq \text{diag} (\wt \sigma_{i,1}^\prime,\wt \sigma_{i,2}^\prime, \ldots, \wt \sigma_{i,M}^\prime)
\end{equation}
where
\begin{align}
\wt \sigma_{i,u}^\prime \triangleq
\begin{cases}
0, \quad \text{for} \; u \leq q,\\
1, \quad \text{for} \; u > q.
\end{cases}
\end{align}
Using above definitions we can now define our projection matrix, i.e.,
\begin{equation}\label{eqn:ProjDefinition}
\mbf P_i \triangleq \mbf V_i \bsymwt \Sigma_i^\prime \mbf V_i^H.
\end{equation}
Below, we show two properties of projection matrices showing that $\mbf P_i$ is a valid projection matrix.

\setcounter{theorem}{0}
\begin{property}
$\mbf P_i \in \mbb C^{M \times M}$ is a projection matrix if and only if $\mbf P_i = \mbf P_i^H = \mbf P_i^2$.
\end{property}
\begin{proof}
Let's start by showing the `only if' part. First, we show $\mbf P_i = \mbf P_i^H$. Taking Harmition of equation \eqref{eqn:ProjDefinition} we have
\begin{equation}\label{eqn:Projhermitian}
\mbf P_i^H = (\mbf V_i \bsymwt \Sigma_i^\prime \mbf V^H)^H = \mbf P_i.
\end{equation}
Now, squaring equation \eqref{eqn:ProjDefinition} we have
\begin{equation}\label{eqn:ProjSquare}
\mbf P_i^2 = \mbf V_i \bsymwt \Sigma_i \mbf V^H \times \mbf V_i \bsymwt \Sigma_i \mbf V^H = \mbf P_i
\end{equation}
where above equation follows from $ \mbf V^H \mbf V_i = \mbf I$ (since they are orthonormal matrices) and $(\bsymwt \Sigma_i^\prime)^2 =\bsymwt \Sigma_i^\prime$ (by construction). From equations \eqref{eqn:Projhermitian} and \eqref{eqn:ProjSquare} it follows that $\mbf P_i = \mbf P_I^H = \mbf P_i^2$.
Next, we show $\mbf P_i$ is a projector by showing that if $\mbf v \in$ range ($\mbf P_i$), then $\mbf P_i \mbf v = \mbf v$, i.e., for some $\mbf w, \mbf v = \mbf P_i \mbf w$, then
\begin{equation}
\mbf P_i \mbf v = \mbf P_i (\mbf P_i \mbf w) = \mbf P_i^2 \mbf w = \mbf P_i \mbf w = \mbf v.
\end{equation}
Moreover, $\mbf P_i \mbf v - \mbf v \in$ null($\mbf P_i$), i.e., 
\begin{equation}
\mbf P_i (\mbf P_i \mbf v - \mbf v) = \mbf P_i^2 \mbf v -\mbf P_i \mbf v = \mbf P_i \mbf v - \mbf P_i \mbf v = \mbf 0.
\end{equation}
This concludes our proof.

\end{proof}

\begin{property}
$\mbf P_i \in \mbb C^{M\times M}$ is an orthogonal projection matrix onto the null space of  $\mbf H_i \in \mbb C^{N^{\text{BS}}\times M}$
\end{property}
\begin{proof}
Since $\mbf P_i = \mbf P_i^H$, we can write 
\begin{equation}
\mbf H_i \mbf P_i^H = \mbf U_i \bsymwt \Sigma_i \mbf V_i^H \times \mbf V_i \bsymwt \Sigma_i^\prime \mbf V^H = \bsym 0.
\end{equation}
The above results follows from noting that $\bsymwt \Sigma_i \bsymwt \Sigma_i^\prime = \bsym 0$ by construction.
\end{proof}

The formation of projection matrix in the waveform design process is presented in the form of Algorithm \ref{alg:Proj}.

%\subsection{Spectrum Sharing and Projection Algorithms}

\begin{algorithm}
\caption{Projection Algorithm}\label{alg:Proj}
\begin{algorithmic}
\IF {$\mbf H_i$ received from waveform design algorithm}
	\STATE{Perform SVD on $\mbf H_i$ (i.e. $\mbf H_i=\mbf U_i \bsym \Sigma_i \mbf V_i^H$)}
	\STATE{Construct $\bsymwt \Sigma_i = \text{diag} (\wt \sigma_{i,1}, \wt \sigma_{i,2}, \ldots, \wt \sigma_{i,p})$}
	\STATE{Construct $\bsymwt {\Sigma}_i^\prime = \text{diag} (\wt \sigma_{i,1}^\prime,\wt \sigma_{i,2}^\prime, \ldots, \wt \sigma_{i,M}^\prime)$}
	\STATE{Setup projection matrix $\mbf P_i = \mbf V_i \bsymwt \Sigma_i^\prime \mbf V_i^H$.}	
\STATE{Send $\mbf P_i$ to waveform design algorithm.}	
\ENDIF 

\end{algorithmic}
\end{algorithm}
%\begin{figure*}[t]

%\end{figure*}

\section{Waveform Design for Spectrum Sharing}\label{sec:nsp}

In the previous section, we designed finite alphabet constant-envelope QPSK waveforms by solving a beampattern matching optimization problem. In this section, we extend the beampattern matching optimization problem and introduce new constraints in order to tailor waveforms that don't cause interference to communication systems when MIMO radar and communication systems are sharing spectrum. We design spectrum sharing waveforms for two cases: the first case is for a stationary maritime MIMO radar and the second case is for moving maritime MIMO radar. The waveform design in these contexts is and its performance is discussed in the next sections.

%We design the MIMO radar waveform with the additional constraint of the waveform being in the null space of the interference channel. In addition, we design the waveform without this constraint but project the designed waveform onto the null space of the interference channels. 

\subsection{Stationary maritime MIMO radar}\label{sec:NSP_included}

\begin{figure}[t]
\centering
\includegraphics[width=3.4in]{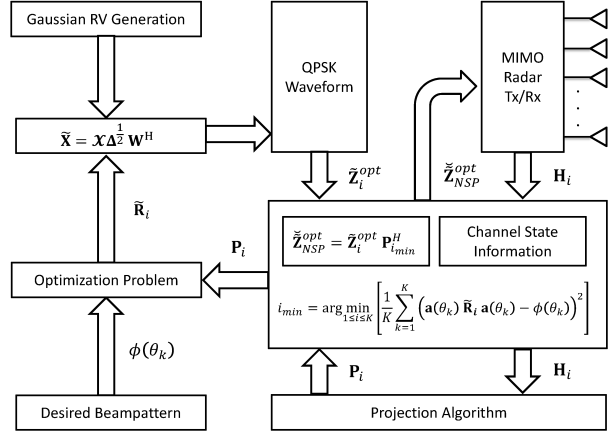}
\caption{Block diagram of waveform generation process for a stationary MIMO radar with spectrum sharing constraints.} 
\label{fig:block_in_opt}
\end{figure}

Consider a naval ship docked at the harbor. The radar mounted on top of that ship is also stationary. The interference channels are also stationary due to non-movement of ship and BSs. In such a scenario, the CSI has little to no variations and thus it is feasible to include the constraint of NSP, equation \eqref{eqn:ProjectedSignal}, into the optimization problem. Thus, the new optimization problem is formulated as 
%equation \eqref{eqn:opt_unct_proj}.
\begin{align}\label{eqn:opt_unct_proj}
\min_{\psi_{ij},\psi_l} &\frac{1}{K} \sum_{k=1}^K \Bigg[ \frac{2}{\pi} \mbf a^H(\theta_k) \mbf P_i \bigg\{\sin^{-1} \bigg(\Re (\mbfwt U)^H \Re (\mbfwt U) + \Im (\mbfwt U)^H \Im (\mbfwt U)\bigg)  + \jmath \sin^{-1} \bigg( \Re (\mbfwt U)^H\Im (\mbfwt U)&-\Im (\mbfwt U)^H\Re (\mbfwt U)\bigg) \bigg\} \nonumber \\
&\times  \mbf P_i^H \mbf a^H(\theta_k) - \alpha \phi(\theta_k) \Bigg]^2\cdot
\end{align}
%\begin{align}\label{eqn:opt_unct_proj}
%&\min_{\psi_{ij},\psi_l} \frac{1}{K} \sum_{k=1}^K \Bigg[ \frac{2}{\pi} \mbf a^H(\theta_k) \mbf P_i \bigg\{\sin^{-1} \bigg(\Re (\mbfwt U)^H \Re (\mbfwt U) \nonumber \\
%&+ \Im (\mbfwt U)^H \Im (\mbfwt U)\bigg)  + \jmath \sin^{-1} \bigg( \Re (\mbfwt U)^H\Im (\mbfwt U)-\Im (\mbfwt U)^H\Re (\mbfwt U)\bigg) \bigg\}  \nonumber \\
%&\mbf P_i^H \mbf a^H(\theta_k) - \alpha \phi(\theta_k) \Bigg]^2
%\end{align}
A drawback of this approach is that it does not guarantee to generate constant-envelope radar waveform. However, the designed waveform is in the null space of the interference channel, thus, satisfying spectrum sharing constraints. The waveform generation process is shown using the block diagram of Figure \ref{fig:block_in_opt}. Note that, $\mc K$ waveforms are designed, as we have $\mc K$ interference channels that are static. Using the projection matrix $\mbf P_i$, the NSP projected waveform can be obtained as
\begin{equation}\label{eqn:ProjectedSignal}
\breve{\mbfwt Z}_{\text{NSP}}^{\text{opt}} = \mbfwt Z_i^{\text{opt}} \mbf P_i^H.
\end{equation}
The correlation matrix of the NSP waveform is given as
\begin{equation}
\breve{\mbfwt R}_i = \frac{1}{N} \left(\breve{\mbfwt Z}_{\text{NSP}}^{\text{opt}}\right)^H  \breve{\mbfwt Z}_{\text{NSP}}^{\text{opt}}  .
\end{equation}
We propose to select the transmitted waveform with covariance matrix $\breve{\mbfwt R}_i$ is as close as possible to the desired covariance matrix, i.e., 
\begin{align}
i_{\text{min}} &\triangleq \argmin_{1 \leq i \leq \mc K} \Bigg[ \frac{1}{K}\sum_{k=1}^{K}{\Big(\mbf{a}^H(\theta_k)\, \breve{\mbfwt{R}}_i\, \mbf{a}(\theta_k)-\phi(\theta_k)\Big)^2} \Bigg]\\
{\mbfwt R}_{\text{NSP}}^{\text{opt}}  &\triangleq \breve{\mbfwt R}_{i_{\text{min}}}.
\end{align}
Equivalently, we select $\mbf P_i$ which projects maximum power at target locations. Thus, for stationary MIMO radar waveform with spectrum sharing constraints we propose Algorithm \eqref{alg:Stationary_Waveform}.

\begin{algorithm}
\caption{Stationary MIMO Radar Waveform Design Algorithm with Spectrum Sharing Constraints}\label{alg:Stationary_Waveform}
\begin{algorithmic}
\LOOP
	\FOR{$i=1:\mc K$}
		\STATE{Get CSI of $\mbf H_i$ through feedback from the $i^{\text{th}}$ BS.}
		\STATE{Send $\mbf H_i$ to Algorithm \eqref{alg:Proj} for the formation of projection matrix $\mbf P_i$.}
		\STATE{Receive the $i^{\text{th}}$ projection matrix $\mbf P_i$ from Algorithm \eqref{alg:Proj}.}
		\STATE{Design QPSK waveform $\mbfwt Z_i^{\text{opt}}$ using the optimization problem in equation \eqref{eqn:opt_unct_proj}.}
		\STATE{Project the QPSK waveform onto the null space of $i^{\text{th}}$ interference channel using $\breve{\mbfwt Z}_{\text{NSP}}^{\text{opt}} = \mbfwt Z_i^{\text{opt}} \mbf P_i^H$.}
	\ENDFOR
	\STATE{Find {$i_{\text{min}} = \argmin\limits_{1 \leq i \leq \mc K} \Bigg[ \dfrac{1}{K}\sum_{k=1}^{K}{\Big(\mbf{a}^H(\theta_k)\, \breve{\mbfwt{R}}_i\, \mbf{a}(\theta_k)-\phi(\theta_k)\Big)^2} \Bigg]$.}}
\STATE{Set ${\mbfwt R}_{\text{NSP}}^{\text{opt}} = \breve{\mbfwt R}_{i_{\text{min}}}$ as the covariance matrix of the desired NSP QPSK waveforms to be transmitted.}
%\STATE{Perform null space projection i.e. $\breve{\mbf x}(t) = \breve{\mbf P} \mbf x(t)$.}
%	\STATE{Receive $V$ from MNSP}
%	\STATE{Project}
\ENDLOOP
\end{algorithmic}
\end{algorithm}

\begin{figure}[t]
\centering
\includegraphics[width=3.4in]{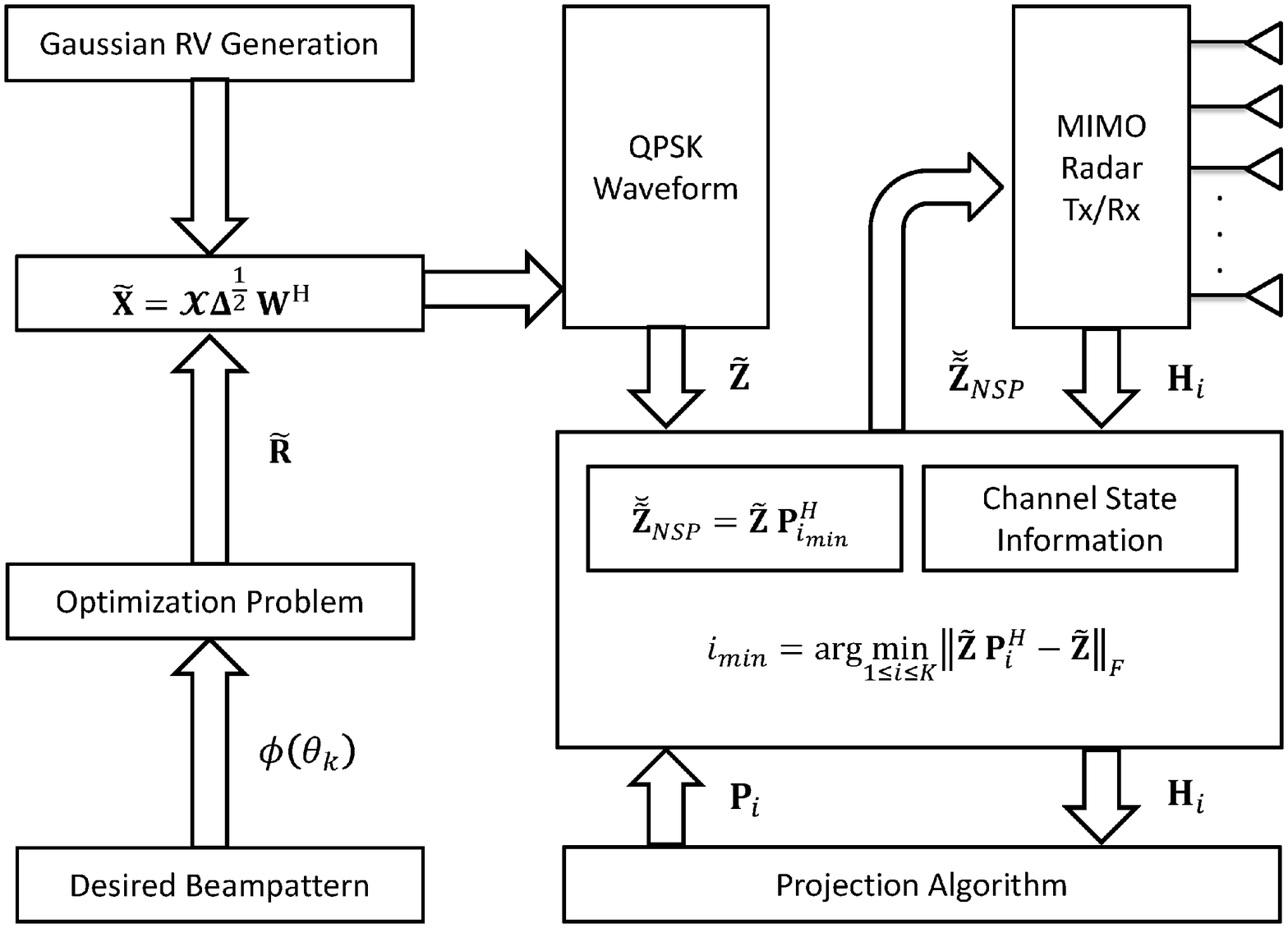}
\caption{Block diagram of waveform generation process for a moving MIMO radar with spectrum sharing constraints.} 
\label{fig:block_out_opt}
\end{figure}

\subsection{Moving maritime MIMO radar}\label{sec:NSP_after}

Consider the case of a moving naval ship. The radar mounted on top of the ship is also moving, thus, the interference channels are varying due to the motion of ship. Due to time-varying ICSI, it is not feasible to include the NSP in the optimization problem. For this case, we first design finite alphabet constant-envelope QPSK waveforms, using the optimization problem in equation \eqref{eqn:JthetaDetailed}, and then use NSP to satisfy spectrum sharing constraints using transform
\begin{equation}\label{eqn:QPSK_NSP_waveform}
\breve{\mbfwt{Z}}_{i} = \mbfwt{Z}\mbf{P}_{i}^H.
\end{equation}
The waveform generation process is shown using the block diagram of Figure \ref{fig:block_out_opt}. Note that only one waveform is designed using the optimization problem in equation \eqref{eqn:JthetaDetailed} but $\mc K$ projection operations are performed via equation \eqref{eqn:QPSK_NSP_waveform}. The transmitted waveform is selected on the basis of minimum Forbenius norm with respect to the designed waveform using the optimization problem in equation \eqref{eqn:JthetaDetailed}, i.e.,
\begin{align}
i_{\text{min}} &\triangleq \argmin_{1 \leq i \leq \mc K} || \mbfwt Z \mbf P_i^H - \mbfwt Z ||_F \\
\breve{\mbfwt Z}_{\text{NSP}} &\triangleq \breve{\mbfwt Z}_{i_{\text{min}}}.
\end{align}
The correlation matrix of this transmitted waveform is given as
\begin{equation}
{\mbfwt R}_{\text{NSP}} = \frac{1}{N} \breve{\mbfwt Z}_{\text{NSP}}^H  \breve{\mbfwt Z}_{\text{NSP}}.
\end{equation}
Thus, for moving MIMO radar waveform with spectrum sharing constraints we propose Algorithm \eqref{alg:Moving_Waveform}.
\begin{algorithm}
\caption{Moving MIMO Radar Waveform Design Algorithm with Spectrum Sharing Constraints}\label{alg:Moving_Waveform}
\begin{algorithmic}
\STATE{Design FACE QPSK waveform $\mbfwt Z$ using the optimization problem in equation \eqref{eqn:JthetaDetailed}.}
\LOOP
	\FOR{$i=1:\mc K$}
		\STATE{Get CSI of $\mbf H_i$ through feedback from the $i^{\text{th}}$ BS.}
		\STATE{Send $\mbf H_i$ to Algorithm \eqref{alg:Proj} for the formation of projection matrix $\mbf P_i$.}
		\STATE{Receive the $i^{\text{th}}$ projection matrix $\mbf P_i$ from Algorithm \eqref{alg:Proj}.}
		\STATE{Project the FACE QPSK waveform onto the null space of $i^{\text{th}}$ interference channel using $\breve{\mbfwt{Z}}_{i} = \mbfwt{Z}\mbf{P}_{i}^H$.}
	\ENDFOR
	
	\STATE{Find {$i_{\text{min}} = \argmin\limits_{1 \leq i \leq \mc K} || \mbfwt Z \mbf P_i^H - \mbfwt Z ||_F$.}}
\STATE{Set ${\mbfwt R}_{\text{NSP}}$ as the covariance matrix of the desired NSP QPSK waveforms to be transmitted.}
%\STATE{Perform null space projection i.e. $\breve{\mbf x}(t) = \breve{\mbf P} \mbf x(t)$.}
%	\STATE{Receive $V$ from MNSP}
%	\STATE{Project}
\ENDLOOP
\end{algorithmic}
\end{algorithm}

%Alternately, we can design CE waveforms by solving the optimization problem in equation \eqref{eqn:opt_unct} and then projection the waveform onto the null space of interference channel. Thus, the CE waveform is generated according to the method explained in Section \ref{sec:gaussian_syn} and then projected onto the null space of the interference channel according to 
%
%This formulation projects the CE waveform onto the null space of the interference channel and we study its impact on the radar waveform performance. 

%\begin{figure*}
%\centering
%%	\includegraphics[width=\linewidth]{../figs/DySPAN_BPSK_outside_opt_width_20.eps} 
%	\caption{Transmit beampattern and its MSE for a \textit{moving} maritime MIMO radar. The figure compares the desired beampattern with the average beampattern of BPSK waveforms for null space projection \textit{after} optimization for candidate interference channel $\mbf H_{\text{Best}}$.}
%		\label{fig:DySPAN_BPSK_outside_opt_width_20}
%\end{figure*}

\section{Simulation}\label{sec:simulation}

In order to design QPSK waveforms with spectrum sharing constraints, we use a uniform linear array (ULA) of ten elements, i.e., $M=10$, with an inter-element spacing of half-wavelength. Each antenna transmits waveform with unit power and $N=100$ symbols. We average the resulting beampattern over 100 Monte-Carlo trials of QPSK waveforms. At each run of Monte Carlo simulation we generate a Rayleigh interference channel with dimensions $N_{\text{BS}} \times M$, calculate its null space, and solve the optimization problem for stationary and moving maritime MIMO radar. 

\subsection{Waveform for stationary radar}
In this section, we design the transmit beampattern for a stationary MIMO radar. The desired beampattern has two main lobes from $-60^\circ$ to $-40^\circ$ and from $40^\circ$ to $60^\circ$. The QPSK transmit beampattern for stationary maritime MIMO radar is obtained by solving the optimization problem in equation \eqref{eqn:opt_unct_proj}. 
%The resulting waveform covariance matrix is given by
%\begin{equation*}
%\mbfwt{R}_{\textrm{NSP}}^\textrm{opt} = \frac{1}{N} \Big(\mbfwt{Z}_{\textrm{NSP}}^\textrm{opt}\Big)^H\mbfwt{Z}_{\textrm{NSP}}^\textrm{opt}.
%\end{equation*}
We give different examples to cover various scenarios involving different number of BSs and different configuration of MIMO antennas at the BSs. We also give one example to demonstrate the efficacy of Algorithms \eqref{alg:Proj} and \eqref{alg:Stationary_Waveform} in BS selection and its impact on the waveform design problem.

\noindent
\textbf{Example 1: Cellular System with five BSs and $\{3,5,7\}$ MIMO antennas and stationary MIMO radar}

\begin{figure}
\centering
\includegraphics[width=3.4in]{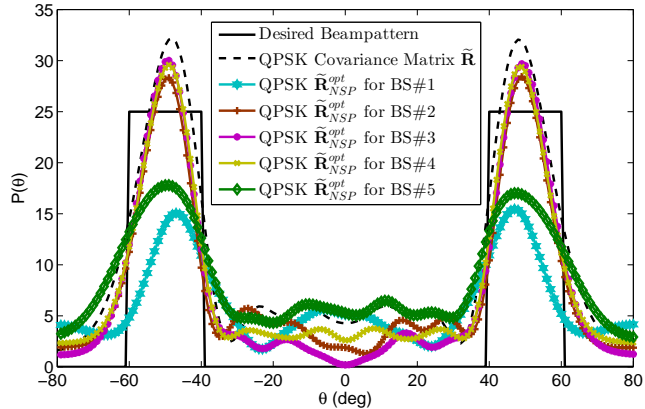} 
	\caption{QPSK waveform for stationary MIMO radar, sharing RF environment with five BSs each equipped with \textit{three} antennas.}
		\label{fig:M10N3_stationary}
\end{figure}

\begin{figure}
\centering
\includegraphics[width=3.4in]{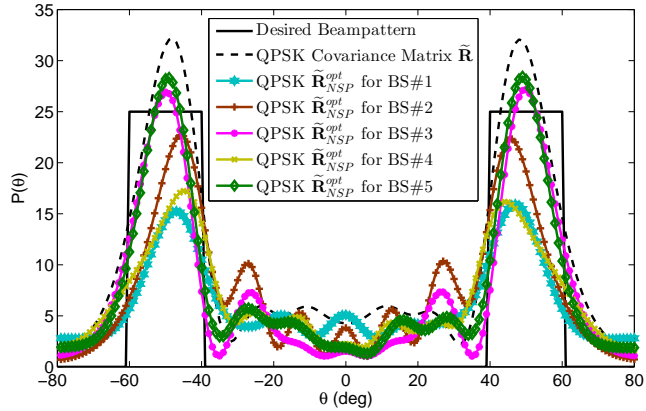} 
	\caption{QPSK waveform for stationary MIMO radar, sharing RF environment with five BSs each equipped with \textit{five} antennas.}
		\label{fig:M10N5_stationary}
\end{figure}

\begin{figure}
\centering
\includegraphics[width=3.4in]{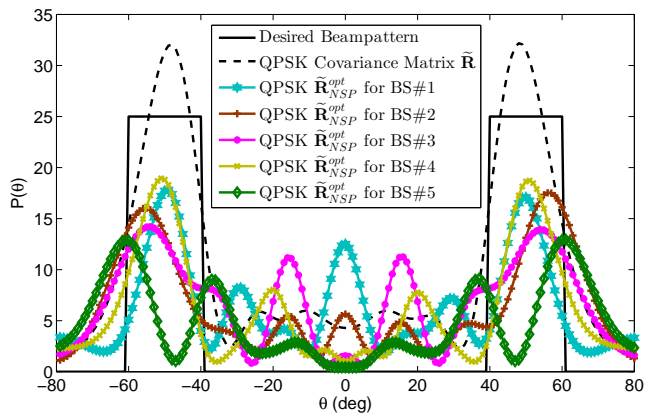} 
	\caption{QPSK waveform for stationary MIMO radar, sharing RF environment with five BSs each equipped with \textit{seven} antennas.}
		\label{fig:M10N7_stationary}
\end{figure}

In this example, we design waveform for a stationary MIMO radar in the presence of a cellular system with five BSs. We look at three cases where we vary the number of BS antennas from $\{3,5,7\}$. In Figure \ref{fig:M10N3_stationary}, we show the designed waveforms for all five BSs each equipped with 3 MIMO antennas. Note that, due to channel variations there is a large variation in the amount of power projected onto target locations for different BSs. But for certain BSs, the projected waveform is close to the desired waveform. In Figure \ref{fig:M10N5_stationary}, we show the designed waveforms for all five BSs each equipped with 5 MIMO antennas. Similar to the previous case, due to channel variations there is a large variation in the amount of power projected onto target locations for different BSs. However, the power projected onto the target is less when compared with the previous case. We increase the number of antennas to $7$ in Figure \ref{fig:M10N7_stationary}, and notice that the amount of power 
projected onto the targets is least as compared to previous two cases. This is because when $N_{\text{BS}} \ll M$ we have a larger null space to project radar waveform and this results in the projected waveform closer to the desired waveform. However, when $N_{\text{BS}} < M$, this is not the case.

%\textbf{Example 2: Cellular System with five BSs each with 5 MIMO antennas}
%In this example, we design waveform for a stationary MIMO radar in the presence of a cellular system with five BSs each equipped with 5 MIMO antennas. In Figure \ref{fig:M10N5_stationary}, we show the designed waveforms for all five BSs. Note that, due to channel variations there is a large variation in the amount of power projected onto target locations for different BSs. 

%\textbf{Example 3: Cellular System with five BSs each with 7 MIMO antennas}
%In this example, we design waveform for a stationary MIMO radar in the presence of a cellular system with five BSs each equipped with 7 MIMO antennas. In Figure \ref{fig:M10N7_stationary}, we show the designed waveforms for all five BSs. Note that, due to channel variations there is a large variation in the amount of power projected onto target locations for different BSs. 

\noindent
\textbf{Example 2: Performance of Algorithms \eqref{alg:Proj} and \eqref{alg:Stationary_Waveform} in BS selection for spectrum sharing with stationary MIMO radar}

\begin{figure}
\centering
\includegraphics[width=3.4in]{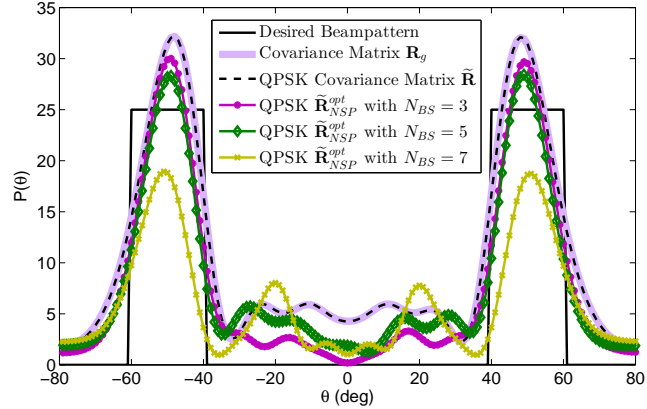} 
	\caption{Algorithm \eqref{alg:Stationary_Waveform} is used to select the waveform which projects maximum power on the targets when $N_{\text{BS}} =\{3,5,7\}$ in the presence of five BSs.}
		\label{fig:QPSK_in_opt}
\end{figure}

In Examples 1, we designed waveforms for different number of BSs with different antenna configurations. As we showed, for some BSs the designed waveform was close to the desired waveform but for other it wasn't and the projected waveform was closer to the desired waveform when $N_{\text{BS}} \ll M$ then when $N_{\text{BS}} < M$. In Figure \ref{fig:QPSK_in_opt}, we use Algorithms \eqref{alg:Proj} and \eqref{alg:Stationary_Waveform} to select the waveform which projects maximum power on the targets or equivalently the projected waveform is closest to the desired waveform. We apply Algorithms \eqref{alg:Proj} and \eqref{alg:Stationary_Waveform} to the cases when $N_{\text{BS}} =\{3,5,7\}$ and select the waveform which projects maximum power on the targets. It can be seen that Algorithm \eqref{alg:Stationary_Waveform} helps us to select waveform for stationary MIMO radar which results in best performance for radar in terms of projected waveform as close as possible to the desired waveform in addition of meeting 
spectrum sharing constraints.

%Different waveforms are generated by changing the number of antennas at the BS and keeping the number of antennas at the MIMO radar fixed. Note that the desired QPSK beampattern and the QPSK beampattern obtained by including the projection matrix inside the optimization problem match closely for $N_{\text{BS}}=3$, i.e, when $N_{\text{BS}} << M$. This is because of a larger null space. It can be observed as the null space becomes smaller the desired waveform and designed waveform are not close enough. In Figure \ref{fig:QPSK_in_opt}, we are illuminating angles from $-60^\circ$ to $-40^\circ$ and from $40^\circ$ to $60^\circ$ with almost 50\% less power when projecting onto the null space of $\mbf H_i$ with dimension $7 \times M$ than $\mbf H_i$ with dimension $3 \times M$. 

\subsection{Waveform for moving radar}
In this section, we design transmit beampattern for a moving MIMO radar. The desired beampattern has two main lobes from $-60^\circ$ to $-40^\circ$ and from $40^\circ$ to $60^\circ$. The QPSK transmit beampattern for moving maritime MIMO radar is obtained by solving the optimization problem in equation \eqref{eqn:35} and then projecting the resulting waveform onto the null space of ${\mbf H_i}$ using the projection matrix in equation \eqref{eqn:QPSK_NSP_waveform}. We give different examples to cover various scenarios involving different number of BSs and different configuration of MIMO antennas at the BSs. We also give one example to demonstrate the efficacy of Algorithms \eqref{alg:Proj} and \eqref{alg:Moving_Waveform} in BS selection and its impact on the waveform design problem.
%The resulting waveform covariance matrix is given by
% and the resulting beampattern is given by equation \eqref{eq:NSPx}.
%\begin{equation*}
%\mbfwt{R}_{\textrm{NSP}} = \frac{1}{N}\mbfwt{Z}^H_\textrm{NSP}\mbfwt{Z}_\textrm{NSP}.
%\end{equation*}

%As with the previous case, different waveforms are generated by changing the number of antennas at the BS and keeping the number of antennas at the MIMO radar fixed. 

\noindent
\textbf{Example 3: Cellular System with five BSs each with $\{3,5,7\}$ MIMO antennas and moving MIMO radar}

\begin{figure}
\centering
\includegraphics[width=3.4in]{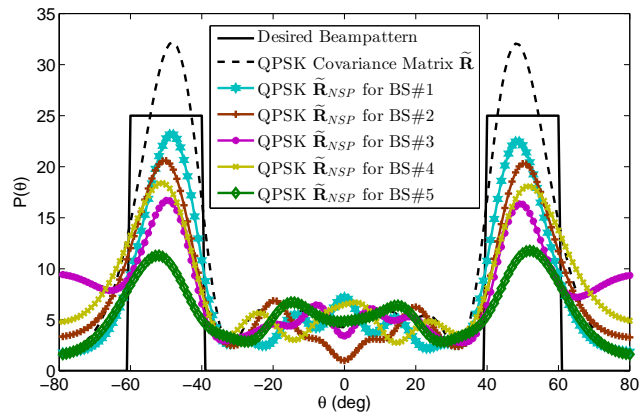} 
	\caption{QPSK waveform for moving MIMO radar, sharing RF environment with five BSs each equipped with \textit{three} antennas.}
		\label{fig:M10N3_mobile}
\end{figure}

\begin{figure}
\centering
\includegraphics[width=3.4in]{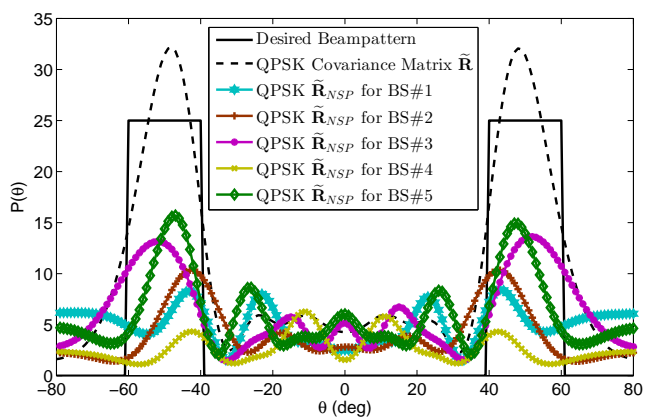} 
	\caption{QPSK waveform for moving MIMO radar, sharing RF environment with five BSs each equipped with \textit{five} antennas.}
		\label{fig:M10N5_mobile}
\end{figure}

\begin{figure}
\centering
\includegraphics[width=3.4in]{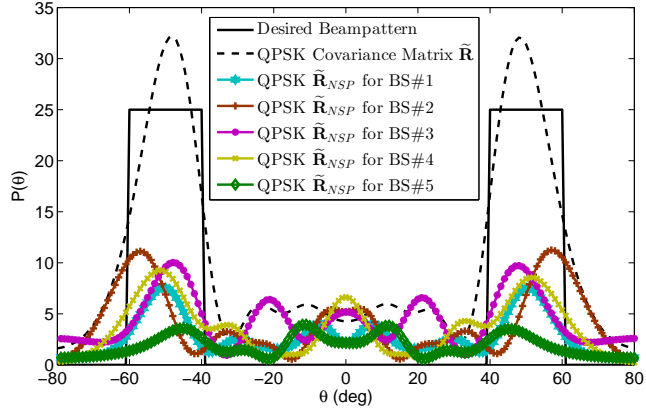} 
	\caption{QPSK waveform for moving MIMO radar, sharing RF environment with five BSs each equipped with \textit{seven} antennas.}
		\label{fig:M10N7_mobile}
\end{figure}

\noindent
In this example, we design waveform for a moving MIMO radar in the presence of a cellular system with five BSs. We look at three cases where we vary the number of BS antennas from $\{3,5,7\}$. In Figure \ref{fig:M10N3_mobile}, we show the designed waveforms for all five BSs each equipped with 3 MIMO antennas. Note that, due to channel variations there is a large variation in the amount of power projected onto target locations for different BSs. When compared with Figure \ref{fig:M10N3_stationary}, the power projected onto the target by NSP waveform is less due to the mobility of radar. In Figure \ref{fig:M10N5_mobile}, we show the designed waveforms for all five BSs each equipped with 5 MIMO antennas. Similar to the previous case, due to channel variations there is a large variation in the amount of power projected onto target locations for different BSs. However, the power projected onto the target is less when compared with the previous case. We increase the number of antennas to $7$ in Figure \ref{fig:M10N7_mobile}, and notice that the amount of power projected onto the targets is least as compared to previous two cases. This is because when $N_{\text{BS}} \ll M$ we have a larger null space to project radar waveform and this results in the projected waveform closer to the desired waveform. However, when $N_{\text{BS}} < M$, this is not the case. Moreover, due to mobility of the radar, the amount of power projected for all three cases considered in this example are less than the similar example considered for stationary radar.

%\textbf{Example 5: Cellular System with five BSs each with 3 MIMO antennas}
%In this example, we design waveform for a stationary MIMO radar in the presence of a cellular system with five BSs each equipped with 7 MIMO antennas. In Figure \ref{fig:M10N2_mobile}, we show the designed waveforms for all five BSs. Note that, due to channel variations there is a large variation in the amount of power projected onto target locations for different BSs. 

%\textbf{Example 6: Cellular System with five BSs each with 5 MIMO antennas}
%In this example, we design waveform for a stationary MIMO radar in the presence of a cellular system with five BSs each equipped with 7 MIMO antennas. In Figure \ref{fig:M10N5_mobile}, we show the designed waveforms for all five BSs. Note that, due to channel variations there is a large variation in the amount of power projected onto target locations for different BSs. 

%\textbf{Example 7: Cellular System with five BSs each with 7 MIMO antennas}
%In this example, we design waveform for a moving MIMO radar in the presence of a cellular system with five BSs each equipped with 7 MIMO antennas. In Figure \ref{fig:M10N7_mobile}, we show the designed waveforms for all five BSs. Note that, due to channel variations there is a large variation in the amount of power projected onto target locations for different BSs. 

\textbf{Example 4: Performance of Algorithms \eqref{alg:Proj} and \eqref{alg:Moving_Waveform} in BS selection for spectrum sharing with moving MIMO radar}

\begin{figure}
\centering
\includegraphics[width=3.4in]{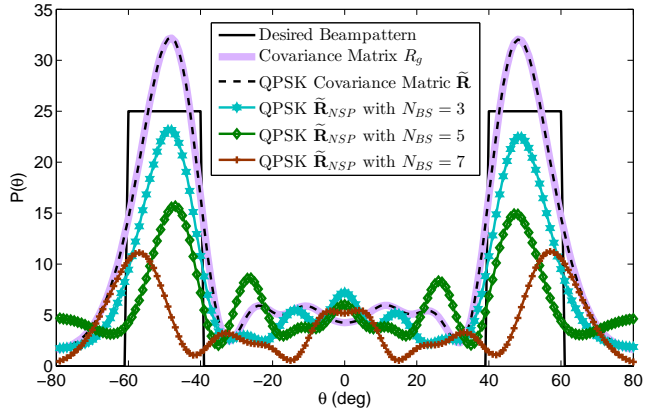} 
	\caption{Algorithm \eqref{alg:Moving_Waveform} is used to select the waveform which projects maximum power on the targets when $N_{\text{BS}} =\{3,5,7\}$ in the presence of five BSs.}
		\label{fig:QPSK_after_opt}
\end{figure}

\noindent
In Examples 3, we designed waveforms for different number of BSs with different antenna configurations. As we showed, for some BSs the designed waveform was close to the desired waveform but for other it wasn't and the projected waveform was closer to the desired waveform when $N_{\text{BS}} \ll M$ then when $N_{\text{BS}} < M$. In Figure \ref{fig:QPSK_after_opt}, we use Algorithms \eqref{alg:Proj} and \eqref{alg:Moving_Waveform} to select the waveform which has the least Forbenius norm with respect to the designed waveform. We apply Algorithms \eqref{alg:Proj} and \eqref{alg:Moving_Waveform} to the cases when $N_{\text{BS}} =\{3,5,7\}$ and select the waveform which has minimum Forbenius norm. It can be seen that Algorithm \eqref{alg:Moving_Waveform} helps us to select waveform for stationary MIMO radar which results in best performance for radar in terms of projected waveform as close as possible to the desired waveform in addition of meeting spectrum sharing constraints.


\section{Conclusion}\label{sec:conclude}
Waveform design for MIMO radar is an active topic of research in the signal processing community. This work addressed the problem of designing MIMO radar waveforms with constant-envelope, which are very desirable from practical perspectives, and waveforms which allow radars to share spectrum with communication systems without causing interference, which are very desirable for spectrum congested RF environments. 

In this paper, we first showed that it is possible to realize finite alphabet constant-envelope quadrature-pulse shift keying (QPSK) MIMO radar waveforms.  
We proved that such the covariance matrix for QPSK waveforms is positive semi-definite and the constrained nonlinear optimization problem can be transformed into an un-constrained nonlinear optimization problem, to realize finite alphabet constant-envelope QPSK waveforms. This result is of importance for both communication and radar waveform designs where constant-envelope is highly desirable.

Second, we addressed the problem of radar waveform design for spectrally congested RF environments where radar and communication systems are sharing the same frequency band. We designed QPSK waveforms with spectrum sharing constraints. The QPSK waveform was shaped in a way that it is in the null space of communication system to avoid interference to communication system. We considered a multi-BS MIMO cellular system and proposed algorithms for the formation of projection matrices and selection of interference channels. We designed waveforms for stationary and moving MIMO radar systems. For stationary MIMO radar we presented an algorithm for waveform design by considering the spectrum sharing constraints. Our algorithm selected the waveform capable to project maximum power at the targets. For moving MIMO radar we presented another algorithm for waveform design by considering spectrum sharing constraints. Our algorithm selected the waveform with the minimum Forbenius norm with respect to the designed waveform. 
This metric helped to select the projected waveform closest to the designed waveform.

\appendices
\section{Preliminaries}
\setcounter{theorem}{0}
This section presents some preliminary results used in the proofs throughout the paper. For proofs of the following theorems, please see the corresponding references. 
\begin{theorem}\label{thm:R_psd}
The matrix $\mbf A \in \mbb{C}^{n \times n}$ is positive semi-definite if and only if $\Re (\mbf A)$ is positive semi-definite \cite{Bern09}. 
\end{theorem}

\begin{theorem}
A necessary and sufficient condition for $\mbf A \in \mbb{C}^{n \times n}$ to be positive definite is that the Hermitian part
\begin{equation*}
 \mbf A_H = \frac{1}{2} \left[ \mbf A + \mbf A^H \right]
 \end{equation*} 
be positive definite \cite{Bern09}. 
\end{theorem}

\begin{theorem}\label{thm:sum_psd}
If $\mbf A \in \mbb{C}^{n \times n}$ and $\mbf B \in \mbb{C}^{n \times n}$ are positive semi-definite matrices then the matrix $\mbf C = \mbf A + \mbf B$ is guaranteed to be positive semi-definite matrix \cite{HJ85}.
\end{theorem}

\begin{theorem}\label{thm:SchurProd}
If the matrix $\mbf A \in \mbb{C}^{n \times n}$ is positive semi-definite then the $p$ times Schur product of $\mbf A$, denoted by $\mbf A_{\circ}^{p}$, will also be positive semi-definite \cite{HJ85}.
\end{theorem}

%
%\begin{lemma}
%
%\end{lemma}

\section{Generating CE QPSK Random Processes From Gaussian Random Variables}\label{sec:ProofGaussian}

Assuming identically distributed Gaussian RV's $\wt x_p, \wt y_p, \wt x_q$ and $\wt y_q$ that are mapped onto QPSK RV's $\wt z_p$ and $\wt z_q$ using
\begin{align}
\wt z_p &= \frac{1}{\sqrt{2}}\Bigg[ \text{sign}\bigg(\frac{\wt x_p}{\sqrt{2}\sigma}\bigg) + \jmath \, \text{sign} \bigg(\frac{\wt y_p}{\sqrt{2}\sigma}\bigg) \bigg]\\
\wt z_q &= \frac{1}{\sqrt{2}}\Bigg[ \text{sign}\bigg(\frac{\wt x_q}{\sqrt{2}\sigma}\bigg) + \jmath \, \text{sign} \bigg(\frac{\wt y_q}{\sqrt{2}\sigma}\bigg) \bigg]
\end{align}
where $\sigma^2$ is the variance of Gaussian RVs. The cross-correlation between QPSK and Gaussian RVs can be derived as
\begin{align}
\mbb E\{\wt z_p \wt z_q^*\} = \frac{1}{2}\mbb E \Bigg[ \bigg\{ &\text{sign}\Big(\frac{\wt x_p}{\sqrt{2}\sigma}\Big) + \jmath \, \text{sign} \Big(\frac{\wt y_p}{\sqrt{2}\sigma}\Big) \bigg\} \bigg\{ \text{sign}\Big(\frac{\wt x_q}{\sqrt{2}\sigma}\Big) + \jmath \, \text{sign} \Big(\frac{\wt y_q}{\sqrt{2}\sigma}\Big) \bigg\} \Bigg]\cdot 
\end{align}
Using equation \eqref{eqn:ExpProperty} we can write the above equation as
\begin{align}\label{eqn:60}
\mbb E\{\wt z_p \wt z_q^*\} = \mbb E \bigg\{&\text{sign}\Big(\frac{\wt x_p}{\sqrt{2}\sigma}\Big) \text{sign} \Big(\frac{\wt x_q}{\sqrt{2}\sigma}\Big) \bigg\} + \jmath \,  \mbb E \bigg\{\text{sign}\Big(\frac{\wt y_p}{\sqrt{2}\sigma}\Big) \text{sign} \Big(\frac{\wt x_q}{\sqrt{2}\sigma}\Big) \bigg\}\cdot
\end{align}
The cross-correlation relationship between Gaussian and QPSK RVs can be derived by first considering
\begin{align}\label{eqn:61}
\mbb E \bigg\{\text{sign}\Big(\frac{\wt x_p}{\sqrt{2}\sigma}\Big)&\text{sign} \Big(\frac{\wt x_q}{\sqrt{2}\sigma}\Big) \bigg\} = \int\limits_{-\infty}^{\infty}\int\limits_{-\infty}^{\infty} \Bigg[\text{sign}\Big(\frac{\wt x_p}{\sqrt{2}\sigma}\Big) \times \text{sign} \Big(\frac{\wt x_q}{\sqrt{2}\sigma}\Big) p (\wt x_p,\wt x_q,\rho_{\wt x_p \wt x_q}) \Bigg] \: d\wt x_p \: d\wt x_q
\end{align}
where $p (\wt x_p,\wt x_q,\rho_{\wt x_p \wt x_q})$ is the joint probability density function of $\wt x_p$ and $\wt x_q$, and $\rho_{\wt x_p \wt x_q} = \frac{\mbb E \{\wt x_p \wt x_q^*\}}{\sigma^2}$ is the cross-correlation coefficient of $\wt x_p$ and $\wt x_q$. Using Hermite polynomials \cite{Bro68}, the above double integral can be transformed as in \cite{ATPM11_FACE}. Thus, 
\begin{align}\label{eqn:63}
\mbb E \bigg\{\text{sign}\Big(\frac{\wt x_p}{\sqrt{2}\sigma}\Big) \text{sign} \Big(\frac{\wt x_q}{\sqrt{2}\sigma}\Big) \bigg\} =& \sum_{n=0}^{\infty} \dfrac{\rho^n_{\wt x_p \wt x_q}}{2\pi\sigma^22^nn!} \times \int\limits_{-\infty}^{\infty} \text{sign}\Big(\frac{\wt x_p}{\sqrt{2}\sigma}\Big)  e^{{\wt x^2_p}/{2\sigma^2}} H_n\Big(\frac{\wt x_p}{\sqrt{2}\sigma}\Big) \: d\wt x_p  
\nonumber \\&\times \int\limits_{-\infty}^{\infty} \text{sign}\Big(\frac{\wt x_q}{\sqrt{2}\sigma}\Big) e^{{\wt x^2_q}/{2\sigma^2}} H_n\Big(\frac{\wt x_q}{\sqrt{2}\sigma}\Big) \: d\wt x_q
\end{align}
where 
\begin{equation}\label{eqn:HermitePoly}
H_n (\wt x_m) = (-1)^n e^{\frac{\wt x^2_m}{2}}\frac{d^n}{d\wt x^n_m}e^{\frac{-\wt x^2_m}{2}}
\end{equation}
is the Hermite polynomial. By substituting $\hat x_p = \frac{\wt x_p}{\sqrt{2}\sigma}$ and $\hat x_q = \frac{\wt x_q}{\sqrt{2}\sigma}$, and splitting the limits of integration into two parts, equation \eqref{eqn:63} can be simplified as
\begin{align}\label{eqn:65}
\mbb E \bigg\{\text{sign}(\hat x_p) \text{sign} (\hat x_q) \bigg\} = \sum_{n=0}^{\infty} \dfrac{\rho^n_{\hat x_p \hat x_q}}{\pi 2^nn!}  \Bigg( \int\limits_{0}^{\infty} e^{\hat x^2_p} \Big[ H_n(\hat x_p)- H_n(-\hat x_p)\Big]  \: d\hat x_p \Bigg)^2\cdot
\end{align} 
Using $H_n(-\hat x_p) = (-1)^n H_n(\hat x_p)$ \cite{DDF10}, equation \eqref{eqn:65} can be written as
\begin{align}
\mbb E \bigg\{\text{sign}(\hat x_p) \text{sign} (\hat x_q) \bigg\} = \sum_{n=0}^{\infty} \dfrac{\rho^n_{\hat x_p \hat x_q}}{\pi 2^nn!}  \Bigg( \int\limits_{0}^{\infty} e^{\hat x^2_p} H_n(\hat x_p) \big(1-(-1)^n\big) \: d\hat x_p \Bigg)^2\cdot
\end{align} 
The above equation is non-zero for odd $n$ only, therefore, we can rewrite it as
\begin{align}\label{eqn:66}
\mbb E \bigg\{\text{sign}(\hat x_p) \text{sign} (\hat x_q) \bigg\} = &\sum_{n=0}^{\infty} \dfrac{\rho^{2n+1}_{\hat x_p \hat x_q}}{\pi 2^{2n}(2n+1)!} \Bigg( \int\limits_{0}^{\infty} e^{\hat x^2_p} H_{2n+1}(\hat x_p) \: d\hat x_p \Bigg)^2 \cdot
\end{align} 
Then using $\int\limits_{0}^{\infty} e^{\hat x^2_p} H_{2n+1}(\hat x_p) \: d\hat x_p = (-1)^n \frac{(2n)!}{n!}$ from \cite{DDF10}, we can write equation \eqref{eqn:66} as
\begin{align}\label{eqn:67}
\mbb E \bigg\{\text{sign}\Big(\frac{\wt x_p}{\sqrt{2}\sigma}\Big) \text{sign} \Big(\frac{\wt x_q}{\sqrt{2}\sigma}\Big) \bigg\} &= \sum_{n=0}^{\infty} \dfrac{\rho^{2n+1}_{\wt x_p \wt x_q}}{\pi 2^{2n}(2n+1)!} 
\Bigg( (-1)^n \dfrac{2n!}{n!}\Bigg)^2 \nonumber \\
&= \dfrac{2}{\pi} \Bigg[ \rho_{\wt x_p \wt x_q} + \dfrac{\rho_{\wt x_p \wt x_q}^3}{2 \cdot 3} + \dfrac{1\cdot3\rho_{\wt x_p \wt x_q}^5}{2 \cdot 4 \cdot 5}+ \dfrac{1\cdot3 \cdot 5\rho_{\wt x_p \wt x_q}^7}{2 \cdot 4 \cdot 6 \cdot 7}+\cdots \Bigg]\nonumber \\
&=  \dfrac{2}{\pi}  \sin^{-1} \bigg(\mbb E \{\wt x_p \wt x_q\} \bigg)
\end{align} 
%%%%%%%%%%%%
In equation \eqref{eqn:61}, we expanded the first part of equation \eqref{eqn:60}. Now, similarly expanding the second part of equation \eqref{eqn:60}, i.e., 
\begin{align}\label{eqn:68}
\mbb E \bigg\{\text{sign}\Big(\frac{\wt y_p}{\sqrt{2}\sigma}\Big)&\text{sign} \Big(\frac{\wt x_q}{\sqrt{2}\sigma}\Big) \bigg\} = \int\limits_{-\infty}^{\infty}\int\limits_{-\infty}^{\infty} \Bigg[\text{sign}\Big(\frac{\wt y_p}{\sqrt{2}\sigma}\Big) &\text{sign} \Big(\frac{\wt x_q}{\sqrt{2}\sigma}\Big) p (\wt y_p,\wt x_q,\rho_{\wt y_p \wt x_q}) \Bigg] \: d\wt y_p \: d\wt x_q
\end{align}
where $p (\wt y_p,\wt x_q,\rho_{\wt y_p \wt x_q})$ is the joint probability density function of $\wt y_p$ and $\wt x_q$, and $\rho_{\wt y_p \wt x_q} = \frac{\mbb E \{\wt y_p \wt x_q^*\}}{\sigma^2}$ is the cross-correlation coefficient of $\wt y_p$ and $\wt x_q$. Using Hermite polynomials, equation \eqref{eqn:HermitePoly}, we can write equation \eqref{eqn:68} as
\begin{align}\label{eqn:69}
\mbb E \bigg\{\text{sign}\Big(\frac{\wt y_p}{\sqrt{2}\sigma}\Big) \text{sign} \Big(\frac{\wt x_q}{\sqrt{2}\sigma}\Big) \bigg\} =& \sum_{n=0}^{\infty} \dfrac{\rho^n_{\wt y_p \wt x_q}}{2\pi\sigma^22^nn!} \times \int\limits_{-\infty}^{\infty} \text{sign}\Big(\frac{\wt y_p}{\sqrt{2}\sigma}\Big) \times e^{{\wt y^2_p}/{2\sigma^2}} H_n\Big(\frac{\wt y_p}{\sqrt{2}\sigma}\Big) \: d\wt y_p  
\nonumber \\
&\times  \int\limits_{-\infty}^{\infty} \text{sign}\Big(\frac{\wt x_q}{\sqrt{2}\sigma}\Big) e^{{\wt x^2_q}/{2\sigma^2}} H_n\Big(\frac{\wt x_q}{\sqrt{2}\sigma}\Big) \: d\wt x_q.
\end{align}
By substituting $\hat y_p = \frac{\wt y_p}{\sqrt{2}\sigma}$ and $\hat x_q = \frac{\wt x_q}{\sqrt{2}\sigma}$, and splitting the limits of integration into two parts, equation \eqref{eqn:69} can be simplified as
\begin{align}\label{eqn:70}
\mbb E \bigg\{\text{sign}(\hat y_p) &\text{sign} (\hat x_q) \bigg\} = \sum_{n=0}^{\infty} \dfrac{\rho^n_{\hat y_p \hat x_q}}{\pi 2^nn!}  \Bigg( \int\limits_{0}^{\infty} e^{\hat y^2_p} \Big[ H_n(\hat y_p)- H_n(-\hat y_p)\Big]  \: d\hat y_p \Bigg)^2\cdot
\end{align} 
Using $H_n(-\hat y_p) = (-1)^n H_n(\hat y_p)$, above equation can be written as
\begin{align}
\mbb E \bigg\{\text{sign}(\hat y_p) \text{sign} &(\hat x_q) \bigg\} = \sum_{n=0}^{\infty} \dfrac{\rho^n_{\hat y_p \hat x_q}}{\pi 2^nn!}  \Bigg( \int\limits_{0}^{\infty} e^{\hat y^2_p} H_n(\hat y_p) \big(1-(-1)^n\big) \: d\hat y_p \Bigg)^2\cdot
\end{align} 
The above equation is non-zero for odd $n$ only, therefore, we can rewrite it as
\begin{align}\label{eqn:72}
\mbb E \bigg\{\text{sign}(\hat y_p) \text{sign} (\hat x_q) \bigg\} = \sum_{n=0}^{\infty} \dfrac{\rho^{2n+1}_{\hat y_p \hat x_q}}{\pi 2^{2n}(2n+1)!} &\Bigg( \int\limits_{0}^{\infty} e^{\hat y^2_p} H_{2n+1}(\hat y_p) \: d\hat y_p \Bigg)^2 \cdot
\end{align} 
Then using $\int\limits_{0}^{\infty} e^{\hat y^2_p} H_{2n+1}(\hat y_p) \: d\hat y_p = (-1)^n \frac{(2n)!}{n!}$, we can write equation \eqref{eqn:72} as
\begin{align}\label{eqn:73}
\mbb E \bigg\{\text{sign}\Big(\frac{\wt y_p}{\sqrt{2}\sigma}\Big) \text{sign} \Big(\frac{\wt x_q}{\sqrt{2}\sigma}\Big) \bigg\} &= \sum_{n=0}^{\infty} \dfrac{\rho^{2n+1}_{\wt y_p \wt x_q}}{\pi 2^{2n}(2n+1)!} 
\Bigg( (-1)^n \dfrac{2n!}{n!}\Bigg)^2 \nonumber \\
&= \dfrac{2}{\pi} \Bigg[ \rho_{\wt y_p \wt x_q} + \dfrac{\rho_{\wt y_p \wt x_q}^3}{2 \cdot 3} + \dfrac{1\cdot3\rho_{\wt y_p \wt x_q}^5}{2 \cdot 4 \cdot 5}+ \dfrac{1\cdot3 \cdot 5\rho_{\wt y_p \wt x_q}^7}{2 \cdot 4 \cdot 6 \cdot 7}+\cdots \Bigg]\nonumber \\
&=  \dfrac{2}{\pi}  \sin^{-1} \bigg(\mbb E \{\wt y_p \wt x_q\} \bigg)\cdot
\end{align} 
Combining equations \eqref{eqn:67} and \eqref{eqn:73}, gives us the cross-correlation of equation \eqref{eqn:60} as
\begin{equation}\label{eqn:ProofXcor}
\mbb E \{\wt z_p \wt z_q\} = \frac{2}{\pi} \Bigg[ \sin^{-1} \bigg(\mbb E \{\wt x_p \wt x_q\} \bigg) + \jmath \, \sin^{-1} \bigg(\mbb E \{\wt y_p \wt x_q\} \bigg) \Bigg]\cdot
\end{equation}

\section{Proofs}\label{app:Proofs}

\begin{proof}[Proof of Lemma \ref{L1}]
To prove Lemma \ref{L1}, we note that the real part of $\mbfwt  R_g $ is $\mbf R_g$ which is positive semi-definite by definition, thus, by Theorem \ref{thm:R_psd}, the complex covariance matrix $\mbfwt  R_g $ is also positive semi-definite.
%\begin{equation}
%\mbfwt  R_{g_H} = \frac{1}{2} \left[ \mbfwt  R_g+  \mbfwt  R_g^H \right] = \Re ( \mbf  R_g  )  = \mbf  R_g 
% \end{equation} 
% Since, $\mbf  R_g $ is positive semidefinite, the Hermitian part $\mbfwt  R_{g_H} $ is also positive definite and thus by Theorem 3 $\mbfwt  R_g$ is positive definite. 
\end{proof}

\begin{proof}[Proof of Lemma \ref{L2}]
To prove Lemma \ref{L2}, we can individually expand the sum, $\sin^{-1} \left( \Re ( \mbfwt R_g ) \right) + \jmath \, \sin^{-1} \left( \Im ( \mbfwt R_g ) \right) $, using Taylor series, i.e., first expanding $\sin^{-1} \left( \Re ( \mbfwt R_g ) \right) $
\begin{align}\label{eq:real_Rg}
\sin^{-1} \left( \Re ( \mbf R_g ) \right) = \Re ( \mbf R_g ) + \frac{1}{2 \cdot 3} \Re ( \mbf R_g )_{\circ}^{3} +  \frac{1 \cdot 3}{2 \cdot 4 \cdot 5} \Re ( \mbf R_g )_{\circ}^{5} + \frac{1 \cdot 3 \cdot 5}{2 \cdot 4 \cdot 6 \cdot 7} \Re ( \mbf R_g )_{\circ}^{7} + \cdots
\end{align}
Then using Theorem \ref{thm:sum_psd}, each term or matrix, on the right hand side, is positive semi-definite, since, $\Re ( \mbf R_g )$ is positive semi-definite by definition. Moreover,  $\sin^{-1} \left( \Re ( \mbf R_g ) \right) $ is also positive semi-definite since its a sum of positive semi-definite matrices, this follows from Theorem \ref{thm:R_psd}.

Similarly, expanding $\jmath \, \sin^{-1} \left( \Im ( \mbf R_g ) \right) $ as
\begin{align}\label{eq:comp_Rg}
\jmath \, \sin^{-1} \left( \Im ( \mbf R_g ) \right) = \jmath [ \Im ( \mbf R_g ) + \frac{1}{2 \cdot 3} \Im ( \mbf R_g )_{\circ}^{3} +  \frac{1 \cdot 3}{2 \cdot 4 \cdot 5} \Im ( \mbf R_g )_{\circ}^{5}  + \frac{1 \cdot 3 \cdot 5}{2 \cdot 4 \cdot 6 \cdot 7} \Im ( \mbf R_g )_{\circ}^{7} + \cdots ]
\end{align}
Now, $\mbfwt R$ is positive semi-definite since real part of it is positive semidefinite, from equation \eqref{eq:real_Rg} and Theorem \ref{thm:SchurProd}.

\end{proof}

\bibliographystyle{ieeetr}
\bibliography{IEEEabrv,QPSK_Proof}
\end{document}